\documentclass[10pt,letterpaper]{article}
\usepackage[top=0.85in,left=0.75in,footskip=0.75in]{geometry}

\usepackage{amsmath}
\usepackage{amssymb}
\usepackage{ulem}

\usepackage{graphicx}
\usepackage{color}

\begin{document}

%\linenumbers

\begin{flushleft}
{\Large
\textbf\newline{Prediction and Comparative Analysis of CTCF Binding Sites based on a First Principle Approach}
}
\newline

Nestor Norio Oiwa\textsuperscript{1,2},
Kunhe Li\textsuperscript{2},
Claudette E. Cordeiro\textsuperscript{3},
Dieter W. Heermann\textsuperscript{2*},

\bigskip
\textbf{1} Department of Basic Science, Universidade Federal Fluminense, Rua Doutor S\'{i}lvio Henrique Braune 22, Centro, 28625-650 Nova Friburgo,  Brazil
\\
\textbf{2} Institute for Theoretical Physics, Heidelberg University, Philosophenweg 19, D-69120 Heidelberg, Germany
\\
\textbf{3} Department of Physics, Universidade Federal Fluminense, Avenida Atl\^antica s/n, Gragoat\'a, 24210-346 Niter\'oi, Brazil
\\
\bigskip

%* nestoroiwa@if.uff.br,nestoroiwa@vm.uff.br
* heermann@tphys.uni-heidelberg.de

\end{flushleft}

\date{\today}

\begin{abstract}

We calculated the patterns for the CCCTC transcription factor (CTCF)
binding sites across many genomes on a first principle approach. 
The validation of the first
principle method was done on the human as well as on the mouse
genome. The predicted human CTCF binding sites are consistent
with the consensus sequence, ChIP-seq data for the K562 cell, nucleosome positions for IMR90 cell as well as the CTCF binding
sites in the mouse HOXA gene. The analysis of Homo sapiens, Mus
musculus, Sus scrofa, Capra hircus and Drosophila melanogaster whole
genomes shows: binding sites are organized in cluster-like groups, where
two consecutive sites obey a power-law with coefficient ranging from to
$0.3292\pm0.0068$ to $0.5409\pm0.0064$; the distance between these
groups varies from $18.08\pm0.52$kbp to $42.1\pm2.0$kbp. 
The genome of Aedes aegypti does not show a power law,
but $19.9\%$ of binding sites are $144\pm4$ and $287\pm5$bp distant of
each other. We run negative tests, confirming the under-representation of
CTCF binding sites in Caenorhabditis elegans, Plasmodium falciparum and
Arabidopsis thaliana complete genomes.
\end{abstract}

%\keywords{CTCF binding sites, Cys2His2 zinc finger, extended ladder model}

%\maketitle

\section{Introduction}

% DNA packing
% ===========

\begin{figure}[t]
   \centering
   \includegraphics[width=0.75\linewidth]{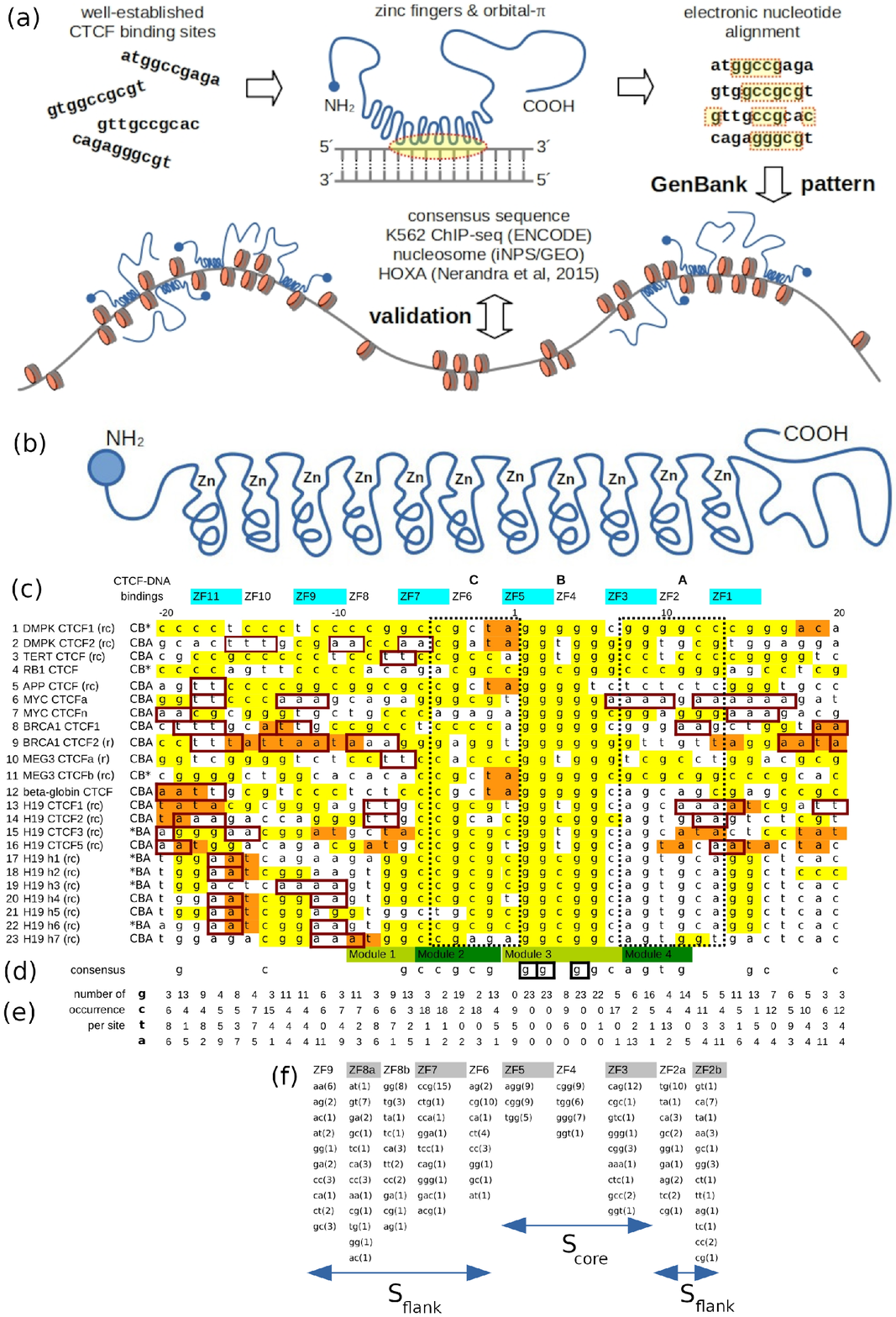}
\caption{ (a) Our workflow. (b) Sketch of the molecular structure of the CCCTC-binding factor (CTCF) with 11 zinc fingers (ZF). Panel (c) shows the electronic nucleotide alignment for CTCF. We indicate the reading direction in reverse and complementary strands with r and c in the parentheses, respectively. Nucleotides with at least 10\% of probability in finding the degenerated ground state  electrons are in  yellow. The highest occupied (HOMO) and the lowest unoccupied molecular orbital (LUMO) nucleotides with at least 10\% of probability in localizing one electron are respectively indicated by orange and red bordered boxes. The eleven CTCF zinc finger positions are marked by the succession of blue and white boxes. The CTCF-DNA binding patterns are indicated by the letters CBA. We indicate respectively the binding positions A and C of ZF2 and ZF6 using dashed boxes. B is about the CTCF-DNA binding of ZF4 and ZF5. We tag respectively by *BA and CB*, when the CTCF miss ZF6 or ZF2 bindings. We remark that there is no nucleotide position zero, in line with \cite{Dreos-2013}.  (d) The consensus sequence is the simple majority (number of alignment nucleotides $\geq$ 12). The black bordered boxes indicate the guanines that appear in all studied CTCF binding sites. The four modulus in \cite{Ong-2014} are indicated in light and dark green. (e) is number of nucleotide occurrence per site. (f) is the nucleotide motifs along the columns with the number of repetitions in parentheses.}
     \label{tabela12}
\end{figure}

In mammals the primary insulator is the CCCTC-binding factor (CTCF), a protein with 10 Cys$_2$His$_2$ and one C$_2$HC zinc finger and the major eukaryotic DNA-protein binding motifs \cite{Klenova-1993,Kim-2007,Xie-2007, Chen-2012} (c.f. Fig. \ref{tabela12}(b)). These transcription factors are characterized  by $3$ to $29$ zinc finger (ZF) units \cite{Iuchi-2001,Ding-2008}, each composed by one zinc ion linking two cysteines at the end of two $\beta$-sheets and two histeines in the C-terminal of one $\alpha$ helix \cite{Wolfe-1999,Klug-2010}. Chromatin immunoprecipitation assays with DNA microarray indicate at least $13,804$ actives binding sites~\cite{Kim-2007} and Xie et al.~\cite{Xie-2007} reports a minimum of $15,000$ binding sites for CTCF, using chromatin immunoprecipitation assay with massively parallel DNA sequencing (ChIP-seq). Chen et al.~\cite{Chen-2012} estimates 326,840 possible sites along the human genome, combining the data from 38 cell lines. Despite CTCF relevance, the quality is poor in 20\% to 30\% of the available data due to limitations of the experimental apparatus and the algorithms for localizing binding site \cite{Kim-2007,Chen-2012,Marinov-2014}. Same mistakes are made, adding false binding sites and making impossible in see the structure of the CTCF distribution. In this paper we present a new method to finding CTCF binding sites based on the interaction of the zinc finger and the electronic cloud of the nucleotide $\pi$-orbital of the double DNA (dDNA) \cite{Oiwa-2016}. We overcome the limitations, unveiling a power law along CTCF biding sites in many complete genomes.

% paper organization
% ==================

The workflow and organization of the article is illustrated in Fig. \ref{tabela12}(a). First of all, we collect 23 experimentally detected CTCF-DNA binding site (see supplementary material S1). Then, we study the electronic cloud of the nucleotide $\pi$-orbital using \cite{Oiwa-2016}. This analysis extends the usual nucleotide alignment based on hydrogen bonds, adding information about the electronic behavior in CTCF binding sites as ground state, higher occupied orbital (HOMO) and lowest unoccupied orbital (LUMO) (see S2). Once we establish a pattern based on our electronic nucleotide alignment, we apply it over a complete genome in multiple genomes (see S3). We validate our putative CTCF binding sites with the consensus sequence \cite{Kim-2007, Bell-2000,  Essien-2009, Ong-2014}, ubiquitous ChIP-seq K562 data \cite{Chen-2012,encode-2007a}, MNase-seq of IMR90 cell with improved nucleosome positioning (iNPS) \cite{GEOmn,Chen-2014,NucMap} and the cluster HOXA \cite{Narendra-2015}. After corroboration of our putative CTCFbs, we study the distribution of CTCFbs over the complete human, mouse, pig, goat, fruit fly and {\it Aedes aegypti} (mosquito) genomes. We use the complete {\it Caenorhabditis elegans, Plasmodium falciparum} and {\it Arabidopsis thaliana} genomes as negative controls. We report cluster-like structures for the CTCF distribution in multiple species. Finally, we discuss the 
limitations of our method as well as ChIP-seq data.

\section{Method}

\subsection{CTCF samples}

% CTCF samples

In order to establish an electronic nucleotide pattern, we consider 23 experimentally confirmed CTCF binding sites, Fig. \ref{tabela12}(b). 
%\sout{, related with 10 different genes} \sout{Bell-2000,Filippova-2001,Renaud-2007,Velazquez-2007, Vostrov-1997, Gombert-2009,Butcher-2004,Wylie-2000,Saitoh-2000,Hark-2000}. 
Detailed descriptions about these CTCFbs are in supplementary material S1. 

The nucleotide sequences in Fig. \ref{tabela12}(b) are fasta or gbk files extracted from the GenBank reference map \cite{GenBank}. We do not use the original sequences from the articles,
%\sout{Bell-2000, Filippova-2001, Renaud-2007, Velazquez-2007, Vostrov-1997, Gombert-2009,Butcher-2004, Wylie-2000, Saitoh-2000, Hark-2000} 
because the literature only publishes the binding site nucleotides. This is insufficient for $\pi$-orbitals. 
We are not restricted just to the nucleotides of the consensus CTCF motif.
The electronic nucleotide descrition of nucleotide $\pi$-orbitals
considers the effects of the surround of the core 20-mers. 
Electrons can easily hop for 16.8 (AT rich sequences) or 25 \AA (CG rich)~\cite{Yoo-2001}, which comprehend at least 5 to 8 bp of the surrounding nucleotides over the core 20-mer binding site. Results with SP1 and EGR1 transcription factors \cite{Oiwa-2016} show the existence of HOMO and LUMO surrounding binding site. Similar phenomena happen for CTCF as we will report in this work, although the biological function of HOMO and LUMO is unknown yet. 
We can easily find the selected binding sites in the GenBank reference map with the same SP1, EGR1, Inr, TATA box and other expected genomics features. All selected binding sites must be experimentally confirmed for multiple methods. 

\subsection{Nucleotide Alignment Using Local Electronic Density of States}

This technique combines DNA melting \cite{Zhu-2007} with the extended ladder model \cite{Senthilkumar-2005, Mehrez-2005}. The  $\pi$-orbital nucleotide along DNA is described as local density of states (LDOS) of the ground state, holes (nucleotides in the valence band without free electron) and higher occupied orbital (HOMO) along with lowest unoccupied orbitals (LUMO), beyond the usual four letters nucleotide alignments \cite{Oiwa-2016}. 

In the context of charge transport, the valence band is the energy levels of the electrons between the ground state and HOMO. The conduction band are the energy levels of the electrons beyond LUMO. Since we have one free electron per nucleotide in the extended ladder model, the valence band will be completely filled and the conduction band will be empty. The ground state electrons are the least mobile, while the HOMO electrons are the most movable ones and they may hop from HOMO to LUMO. In this work, the nucleotides with ground state electrons, marked in yellow in Fig. \ref{tabela12}(c), are actually the nucleotides with at least 10\% of probability of finding the degenerated ground state electrons. The difference between HOMO and LUMO is absent in conductors, while electric insulators present wide gaps. The gap in the extended ladder model \cite{Oiwa-2016} gives a semi-conductor characteristic for the double helix. The computation of LDOS is detailed in the appendix S2.

We perform an electronic alignment instead of traditional nucleotide alignment, considering simultaneously  adenine (A), citosine (C), guanine (G), timine (T), ground states (yellow), HOMO(orange) and LUMO (red bordered boxes), Fig. \ref{tabela12}(c). The final alignment in Fig. \ref{tabela12}(c) is made manually. 

We cannot ignore the symmetries of the genetic code, since CTCF read dDNA in four directions in function of complementary and reflection symmetries. So, the charge patterns of the tips of ZF and the LDOS of DNA chains must be evaluated in the direct or positive strand and direct reading (from 5' to 3'), in the direct strand and reverse reading (from 3' to 5'), complementary or negative strand with direct reading and complementary strand with reverse reading.

\subsection{Pattern identification}

We divide the prediction technique in two parts. In the first part of the technique, we scan the contiguous sequences (contigs), looking for the electronic distribution described in the previous section. Then, we consider the number of nucleotide occurrence and the motifs in Fig. \ref{tabela12}(e,f). Since the length and the number of the binding sites is small in Fig. \ref{tabela12}(c), we do not use any algorithm for motifs detection and classification. We arrange the nucleotides manually. Indeed, there are only four and three observed motifs in the ZF4 and ZF5 triplets, Fig. \ref{tabela12}(f). The number of motifs is reduced in the middle of the binding site, but large in the flanking region. So, we divide the nucleotides in two sets: $S_{\mbox{\scriptsize core}}$ and $S_{\mbox{\scriptsize flank}}$.

In the core of the CTCFBs, we define the geometric average probability $P_{\mbox{\scriptsize core}}(S_{\mbox{\scriptsize core}})= \left[\prod_k P(S_k)\right]^{1/3}$ where $S_{\mbox{\scriptsize core}}=\cup_k S_k$, $k$=\{ZF3, ZF4, ZF5\}, and $P(S_k)$ is the probability of occurrence of the motif $S_k$, Fig. \ref{tabela12}(f). We have a cubic root in $P_{\mbox{\scriptsize core}}$, because we are analyzing the patterns of 3 zinc fingers. After extensive tests localizing the listed Fig. \ref{tabela12}(c) in GenBank flat files, we conclude that a minimum of 9.0\% for $P_{\mbox{\scriptsize core}}$ is required for a valid DNA-CTCF binding.

In the region flanking the core, we define a probability  
$P_{\mbox{\scriptsize flank}}(S_{\mbox{\scriptsize flank}})
         =\frac{1}{2}\left[\prod_kP(S_k)\right]^{1/7}+\frac{1}{2}\left[\prod_iP(S_i)\right]^{1/15}$
where $S_{\mbox{\scriptsize flank}}=\cup_k S_k$, $k$=\{ZF2a, ZF2b, ZF6, ZF7, ZF8a, ZF8b, ZF9\}, $P(S_k)$ is the  probability of occurrence of the motif $S_k$,  and $P(S_i)$ is the probability of the nucleotide occurrence $S_i$ in the position $i$, $i$=-11, ..., -1,10,...,13. The first term $\left[\prod_kP(S_k)\right]^{1/7}$ in $P_{\mbox{\scriptsize flank}}$ guarantees the detection of nucleotide sequences listed in Fig. \ref{tabela12}(f), and we have 7th root in the expression since we are considering 7 elements in $S_k$. However, there are considerable variation in $S_{\mbox{\scriptsize flank}}$, comparing with $P_{\mbox{\scriptsize core}}$. If we restrict the motifs just in Fig. \ref{tabela12}(f), we will miss valid CTCFbs. So, we introduce $\left[\prod_i P(S_i)\right]^{1/15}$ in $P_{\mbox{\scriptsize flank}}$. We decompose the flanking sequence in their 15 nucleotides, $S_i=\{a,t,c,g\}$. 
Then, we estimate the geometric average probability associated with the occurrence of each particular nucleotide $S_i$ along the binding site, Fig. \ref{tabela12}(e). Our tests show that
the probability of a valid CTCFbs $P_{\mbox{\scriptsize flank}}$ should be bigger than 6.5\%.

We illustrate the procedure in the supplementary material S4.

\begin{table}
   \begin{center}
      \begin{tabular}{lrccccc} \hline \hline 
  & $L$(bp) &&$n_{\mbox{\scriptsize ctcf}}$ &$<l_{\mbox{\scriptsize ctcf}}>$ (kbp)& $\alpha$    &$\lambda$ (kbp)\\ \hline

human all*     &2,814,809,546& &331,668& 8.8$\pm$3.1&0.511$\pm$0.014&19.28$\pm$0.24\\
human centromer&   76,305,151& &  1,892&  38$\pm$17 &  \multicolumn{2}{c}{no structure} \\ 
human variable &   14,059,087& &  1,528&  9.207\dag & \multicolumn{2}{c}{no structure} \\ \hline
mouse all\ddag  &2,541,456,020 & &277,027  & 9.4$\pm$1.9& 0.3292$\pm$0.0068&19.79$\pm$0.31 \\ 
mouse chromY &82,248,315 & &2,512&32.742\dag&  \multicolumn{2}{c}{detailed in the text $^a$ }\\
\hline
pig   &2,389,924,585 & &316,919& 7.9$\pm$2.7  &0.484$\pm$0.013 &22.44$\pm$0.37 \\
goat   & 2,462,599,335&  &264,286 &9.7$\pm$3.2 &  0.5409$\pm$0.0064  &24.24$\pm$0.35 \\
fruit fly  &128,506,876 & &8,962&14.4$\pm$2.2  &  0.454$\pm$0.012 &18.08$\pm$0.52\\
fruit fly chrom4& 1,200,662& &20&60.033\dag& \multicolumn{2}{c}{no structure}\\
{\it A. aegypti}  & 1,195,030,408 && 39,777 &  30.043\dag& 144$\pm$4bp,  287$\pm$5bp.$^a$  &42.1$\pm$2.0 \\
{\it C. elegans}  &100,272,607 && 2,086& 48.2$\pm$7.7 &  \multicolumn{2}{c}{no structure}   \\
{\it P. falciparum } &23,264,338 & &46 &  530$\pm$340&\multicolumn{2}{c}{no structure} \\
{\it A. thaliana}  & 116,129,212& &1,595& 72.7$\pm$3.6 & \multicolumn{2}{c}{no structure}  \\
\hline \hline
      \end{tabular} 
   \end{center}

\caption{$L$ is the genome length,
$n_{\mbox{\scriptsize ctcf}}$ is the predicted CTCF binding sites (CTCFbs) and 
$ <l_{\mbox{\scriptsize ctcf}}>$ is the chromosomal average of CTCFbs density. The probability  distribution $P(\Delta)$ of the difference $\Delta$ between two consecutive CTCFbs obeys a scaling law $\alpha$ from 11bp $\leq \Delta\leq$ 2,000bp to 16bp $\leq \Delta\leq$ 17,000bp, depending of the considered genome. $P(\Delta)$ follows an exponential decay with typical length $\lambda$, when we consider the fitting regions from 2,000bp$\leq\Delta\leq$ 78kbp to 9.7kbp$\leq\Delta\leq$ 99kbp. *Heterochromatins were excluded.  \dag No standard deviations due to the reduced amount of data. \ddag Chromosome Y is excluded. $^a$Well-defined $\Delta$ CTCFbs distances.} 
   \label{gbands2}
\end{table}

\section{Validation}

\subsection{Consensus sequence}

The most striking feature of the alignment of 23 CTCFbs in Fig. \ref{tabela12}(c) is the guanine at the positions 2, 3 and 5, marked with a black box in Fig. \ref{tabela12}(d). Actually, guanines at the position 2 and 5 coincide with the middle nucleotide of the triplet of the ZF4 and ZF5 and the amino acid of tip of these ZF tips are base. So, the positive charged tips of ZF4 and ZF5 bind with the ground state electrons of guanines in position 2 and 5; a similar mechanism is described in \cite{Iuchi-2001,Wolfe-1999, Klug-2010, Miller-1985,Nolte-1998}. Coarse-grained Monte Carlo simulations confirm this finding \cite{Lei-2015}. Further, Kim et. al.~\cite{Kim-2007} increases the specificity of their CTCF binding site prediction using these same nucleotides in positions 2 and 5 as well as -4 and 7 (positions 6, 11, 14 and 16 in their article). There is always adsorption of the zinc fingers 4 and 5 by the DNA. 

We do not observe HOMO between -5 to -2 and 2 to 9, and there is no LUMO between -4 to 6. We never observe over-position between ground state and HOMO or LUMO electrons. The core of CTCF-DNA binding sites is a region without mobile electrons and CTCF anchors their zinc fingers in the most stable electrons, i.e. ground state electrons. 

Since the electronic alignment considers the charges in the tips of the zinc-finger \cite{Oiwa-2016}, the eleven ZFs in CTCF reveal more details about the protein-DNA attachment. There are five ZF with well-defined charge motifs:  ZF2, ZF4 ZF5, ZF6 and ZF9. The finger tip is acid (negative) for ZF2 and ZF6 as well as base (positive) for ZF4, ZF5 and ZF9. Electrons in the nucleotides will bind the positive tips, and holes in negative ones. We will ignore ZF9, because it is neither in the core binding site nor fundamental for CTCF-DNA binding \cite{Lei-2015}. ZF4 and ZF5 always bind with the dDNA \cite{Lei-2015}. We do not find any particular property for ZF3. Thus, we will focus on the binding sites for ZF2 and ZF6 (respectively A and C in Fig. \ref{tabela12}(b)) and ZF4 and ZF5 marked as B in Fig. \ref{tabela12}(c). Instead of three nucleotides of the triplet, we consider five nucleotides in A and C, blue box in Fig. \ref{tabela12}(c), because the CTCF is a flexible molecule and the finger may displace back and forward along the double helix. The site B is the triplets under ZF4 and ZF5.  CTCF sometimes misses the binding sites A or C, but it always binds in B.  CTCF-DNA binding is successful only if we do not miss A and C sites simultaneously, Fig.~\ref{tabela12}(c).

The consensus sequence in Fig. \ref{tabela12}(d) is just the simple majority (number of alignment nucleotides $\geq$ 12). We avoid the Schneider and Stephens logo, and we use neither the Shannon Information content, Gibbs binding free energy nor Position Weight Matrix for the calculus of the specific-binding free energy \cite{Schneider-1986,Schneider-1990,Bailey-1994,haeseleer-2006}, because we have neither a clear boundary for the binding for the background sequences nor consider the flanking sequences. We get better results circumventing the intricate heuristic weighting factors and scores of the nucleotide alignments or misalignments \cite{Setubal-1997},  neural networks \cite{Mount-2004}, and we do not use MNase-seq~\cite{Zhong-2014} and ChIP-seq sequences from ENCODE in order to find the motif behind CTCF \cite{encode-2007a,encode-2007b}, simplifying the localization process and saving computational time. Despite the over simplification, we have a good matching with the consensus 5'-ccgcgnggnggcag-3' \cite{Bell-2000,Kim-2007,Essien-2009,Ong-2014}. These nucleotides are divided in four moduli \cite{Ong-2014}. We use the border between module 2 and 3 as the position of reference. So, the nucleotide in position 1 is at the beginning of the module 3. The nucleotide at the position -1 is the first one before nucleotide in position 1. Following the literature, there is no position zero \cite{Dreos-2013}. The modulus 2 in \cite{Ong-2014} is related with ZF6, modulus 3 with ZF4 and ZF5 and modulus 4 with ZF2 and ZF3. ZF9 is maybe connected with modulus 1, but the sequence is at the right of ZF9 triplet and the evidence of consensus sequence is too faint for conclusions \cite{Kim-2007,Ong-2014}.

\subsection{CTCF and ChIP-seq K562 data}

Once we identify the electronic nucleotide pattern and establish a criteria for CTCF binding sites, we localize all human CTCFbs along the assembly hg38, Table \ref{gbands2}. We find 335,088 binding sites. This number is remarkable close to the total cumulative number of 326,840 CTCF binding sites identified by Chen et al. using data from 38 human cell lines \cite{Chen-2012}.

We compare our predicted CTCFbs to the ChIP-seq K562 ubiquitous binding sites. The 8,771 ubiquitous CTCFbs from 5 ENCODE K562 files are described in supplementary material S5. We have  29.8$\pm$3.8\% of perfect match between our method against experimental data. The median Q2 of the distances between  predicted and observed binding sites shows us that 50\% of the putative are just at a 473bp distant from the expected one and 75\% of them (third quartile, Q3) are at the maximum 2,352bp. Beyond Q3, we have some huge discrepancies reaching 73,250bp. As we lay out in the discussion, the discrepancy of the last quartile (25\% of data) between our putative CTCF binding sites and those detected by ChIP-seq comes from the limitations of the chromatin immunoprecipitation technique.

We can improve the matching in light of the helical geometry of the dDNA. When we observe the three-dimensional structure of dDNA, there are two possible grooves where the zinc finger will insert into the dDNA to read the $\pi$-orbital. The major groove is 22 \AA \hspace{0.1cm} large, while the minor groove has only 11 \AA \cite{Wing-1980}. We expect more CTCFbs in the direct strand and direct reading (from 5' to 3') and in complementary strand and reverse reading (from 3' to 5'), since it is easier for the CTCF to insert into the major groove. We can see in Fig.\ref{tabela12}(c) that we have 21 samples in  the major groove and the matching between predicted and ChIP-seq K562 data increases: 34\% of binding sites will have a perfect matching, with Q2=401bp, Q3=2,238bp and a maximum discrepancy of 60,676. However, we have only 22\% of matching, Q2=621bp, Q3=2,348 and a maximum of 73,246 bp difference for CTCF binding in the minor dDNA groove. Here, we linked the minor groove with the direct reading in the complementary strand and reverse reading in the direct strand. In Fig.\ref{tabela12}(c) BRCA1 CTCF2 and MEG3 CTCFa are in the direct strand and reverse reading, associated with the minor glove. The absence of major and minor groove distinction in our method is obvious when we see the chromosomal average proportions of each reading direction: direct strand and direct reading is 29$\pm$1\% of the predicted CTCFbs; direct strand and reverse reading has 21$\pm$1\%; complementary strand and direct reading values 20.8$\pm$0.7\%; and complementary strand and reverse reading is 29.4$\pm$0.7\%. The number of  direct strand and reverse reading as well as complementary strand and direct reading could be overestimated.

\subsection{CTCF and nucleosome}

In order to evaluate the coherence of our findings, we study the nucleosome distribution around our putative binding sites. The nucleosome binding sites are localized using an improved nucleosome positioning algorithm (iNPS) over the sample GSM1095279 from Gene Expression Omnibus database, a MNase-seq assay in human IMR90 fetal lung fibroblast cell \cite{GEOmn, NucMap}. iNPS increases the number of detected nucleosomes \cite{Chen-2014}. A detailed description of iNPS can be found in supplementary material S6. We extract 5,968,503 nucleosomes from this sample, covering 658,606,003bp, resulting in an average nucleosome density $\rho$ of 21\% for human genome. 

\begin{figure}[t]
   \centering
   \includegraphics[width=0.75\linewidth]{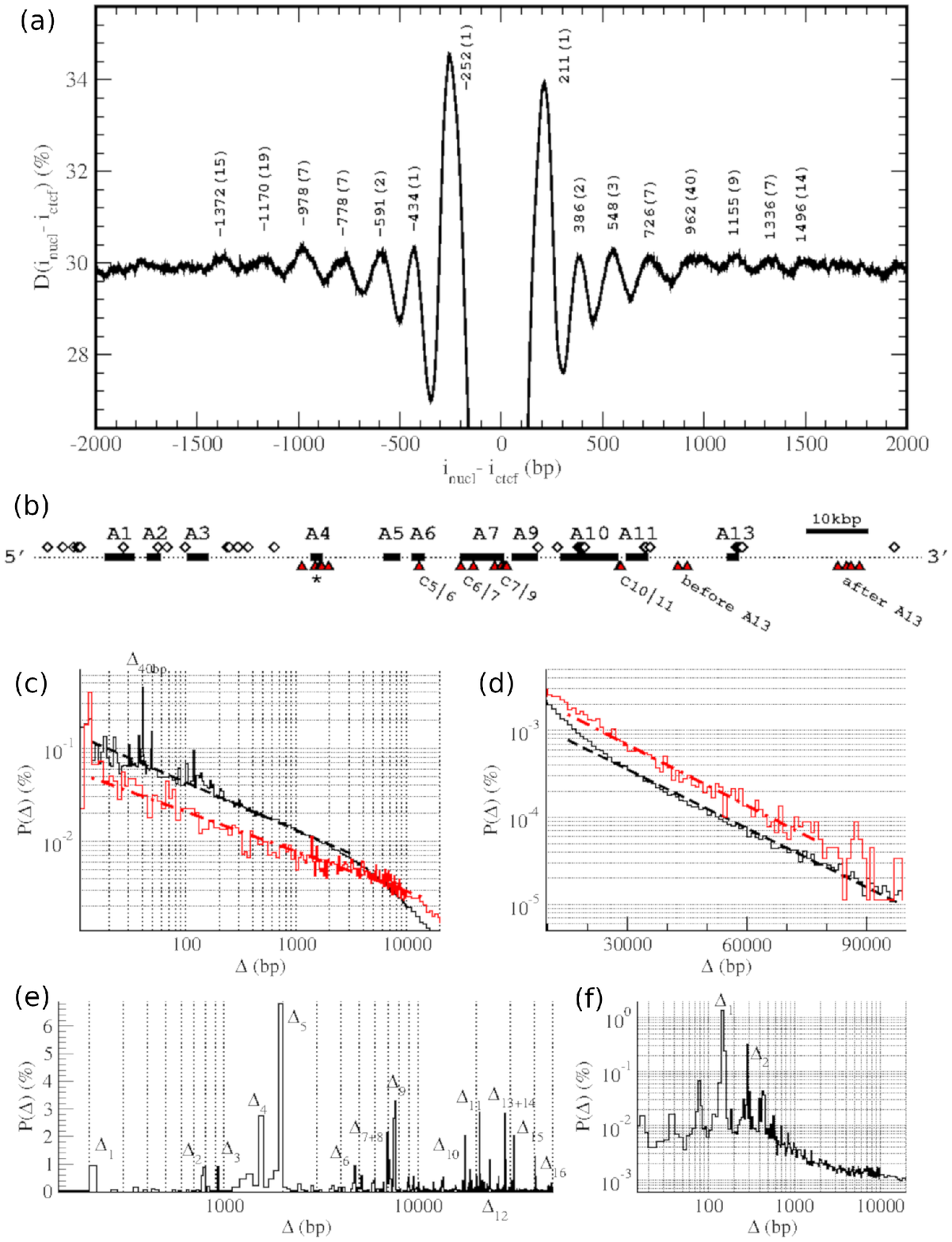}
   \caption{(a) The black line shows the average chromosomal density $\rho(i_{\mbox{\scriptsize nucl}}-i_{\mbox{\scriptsize ctcf}})$ of nucleosomes per nucleotide around the predicted CTCF binding site (CTCFbs) $i_{\mbox{\scriptsize ctcf}}$, excluding the nucleosome in the CTCF position. The exact neighbour nucleosome positions are indicated by numbers above the peaks with the error in the parenthesis. (b) shows the HOXA genes (black line), the predicted (white diamond) and predicted CTCFbs which are experimentally confirmed (red triangle) from the mouse HOXA gene cluster. The red triangles in C5$\vert$6, C6$\vert$7, C7$\vert$9, C10$\vert$11, before and after C13 are the CTCFbs reported in \cite{Narendra-2015}. The asterisk indicates the faint response for CTCF in \cite{Narendra-2015}, not reported by the authors.(c,d) are probabilities $P(\Delta)$ in finding the next consecutive putative CTCF binding site in percentage  against the distance $\Delta$ in base pairs for human (black) and fruit fly (red). (c) The dark dashed and the red dotted dashed lines indicate the power-law for human and {\it D. melanogaster,} respectively. (d) The exponential fitting for human and fly in semi-log scale are also pointed by dark dashed and red dotted dashed lines. (e) $P(\Delta)$ of mouse Chromosome Y with multiple typical $\Delta_1-\Delta_{16}$ distances. (f) {\it Aedes aegypti} $P(\Delta)$ with the characteristic 144$\pm$4bp ($\Delta_1$) and 287$\pm$5bp ($\Delta_2$) distances. }
   \label{fittings01}
\end{figure}

We combine iNPS and our CTCFbs data in Fig. \ref{fittings01}(a). $\rho(i_{\mbox{\scriptsize nucl}}-i_{\mbox{\scriptsize ctcf}})$ is the nucleosome density around CTCFbs for the complete human genome, the main variable $i_{nucl}-i_{ctcf}$ is the nucleosome position minus the position of CTCF center in the unit of bp. In order to improve the quality of the nucleosome peaks, we consider only the CTCFbs of the major DNA groove, because they are less affected by the helical geometry of dDNA as we discussed in the previous section. The average nucleosome density $\rho$ around CTCF binding site is 30\% instead of 21\%, mentioned in the last paragraph. $\rho$ is always higher in CTCFbs rich domain. When we look for nucleosome fluctuation  around each CTCFbs, we find 7 nucleosomes peaks around CTCF in direction of the N-terminus and 8 nucleosomes in the C-terminus direction, while \cite{Chen-2012, Fu-2008} report 20 nucleosomes around CTCFbs. \cite{Chen-2012} uses data for nucleosome and CTCF from the same source, GENCODE \cite{GENCODE}. \cite{Fu-2008} uses CTCFbs from UCSC genome browser \cite{UCSCbrowser} and the nucleosome positions are predictions using \cite{Segal-2006}. Since our analysis is done on a genome-wide scale, the relatively remote nucleosomes are considered to be more fluctuated and hard to recognized in the figure; on the other hand, this also demonstrates the prominence of the observed peaks. Each nucleosome in our work includes 185$\pm$21bp, equivalent to the sum of 147bp necessary to wrap one nucleosome and 38bp for the linker in agreement with \cite{Chen-2012,  Chen-2014,Fu-2008}. \cite{Chen-2012,Chen-2014} describe a symmetrical distribution, because they do not consider the CTCF reading direction. Since the reading direction is available in our analysis, we observe asymmetry in the nucleosome positioning around CTCFbs as \cite{Fu-2008,Clarkson-2019}. The distance between the CTCFbs and the first peak for the C-terminus is shorter than N-terminus as reported in \cite{Clarkson-2019}, and we have a substantial fluctuation in the position 962bp with an error of 40bp, Fig. \ref{fittings01}(a).  We should expect this asymmetry in the nucleosome distribution around CTCFbs, because N and C terminus have different structures.

\subsection{CTCF and HOXA}

Once we have established a reliable protocol for the human genome, we have applied the method to the mouse genome. We find 279,539 CTCF binding sites in the build GRCm38. The whole genomic chromosomal average predicted binding sites $<l_{\mbox{\scriptsize ctcf}}>$ for mouse are comparable with the human genome. 

We also assess our method for the mouse HOXA gene cluster, composed by 11 genes (A1-7, A9-11 and A13), Fig. \ref{fittings01}(b), in order to reproduce Narendra et al. findings \cite{Narendra-2015}. We detect all CTCFbs between A5 and A6 (C5$\vert$6), A6 and A7 (C6$\vert$7), A7 and A9 (C7$\vert$9), A10 and A11 (C10$\vert$11), before and after A13 reported by \cite{Narendra-2015}. Nevertheless, we should evaluate this statement carefully, since our program detects actually 52 binding sites, while \cite{Narendra-2015} reports just 6. There are many CTCFbs organized in cluster-like groups \cite{Chen-2012} as C6$\vert$7 with two binding sites, C7$\vert$9 with 4, C10$\vert$11 with two, the CTCFbs before and after A13 with two and four respectively, Fig.\ref{fittings01}(b). The CTCF assays by \cite{Narendra-2015} have a precision around 1 kbp and are unable to find one particular 20bp long CTCF binding site. In the case of the CTCFbs after A13 gene, the CTCFbs cluster stretches for 3,625bp. The motif based methods in CTCF assays do not consider the local repeats as alternative sites. They choose one of possible sites that may spread for few kbps. Interestingly, the faint signal between the gene A4 and A5, not reported by \cite{Narendra-2015}, is positioned in one putative CTCFbs group, detected by our program. This faint signal is a cluster with seven CTCFbs, asterisk in Fig. \ref{fittings01}(b). We also observe, respectively, a mismatch of 2,981 and 2,795bp between our results and \cite{Narendra-2015} for C5$\vert$6 and C6$\vert$7. The source for this displacement is the considered CTCF consensus motifs. In order to test the robustness of our electronic alignment, we do not include their CTCF motifs (supplemental material in \cite{Narendra-2015}) in our 23 samples, Fig. \ref{tabela12}(b).

In the previous section, we show that we get better results considering only CTCFbs in the major dDNA groove. However, the CTCFbs in the minor DNA cannot be completely neglected. since we have many CTCFbs in Fig. \ref{fittings01}(b). For example, the CTCFbs of C5$\vert$6, C6$\vert$7 and C10$\vert$11 are all in the minor dDNA groove. 

The mismatches between our results and~\cite{Narendra-2015} around C5$\vert$6 and C6$\vert$7 give us an idea about the inaccuracy in the positioning $i_{\mbox{\scriptsize ctcf}}$ of our method. The estimate of misplacement is around 3kbp. However, the most evident feature in  Fig.~\ref{fittings01}(b) is the coalescence of the CTCFbs, reported by \cite{Chen-2012} as clusters of binding sites. However, the concept of cluster demand a gaussian among CTCFbs distribution and we do not observe such structure. Since our electronic alignment is not limited by poor quality data \cite{Marinov-2014,Park-2009} or absence of the expected 20-mer consensus motif \cite{Chen-2012}, we make more accurate analysis.

\section{Results}

%\subsection{Cluster-like structure}

Instead of a cluster organization for CTCFbs suggested by \cite{Chen-2012,Essien-2009}, we implement another evaluation, detecting a power law in $P(\Delta)$, Table \ref{gbands2}  and Fig. \ref{fittings01}(c), indicating organized structure for CTCFbs. Here, $\Delta$ is the distance of two consecutive CTCFbs and $P(\Delta)$ is the probability of finding the next binding site. In humans we adjust $\alpha$ in $P(\Delta)\approx\Delta^\alpha$, considering two or three orders of magnitude.The human euchromatic regions have $\alpha=0.511\pm 0.014$, fitting within the interval 14bp $\leq\Delta \leq$ 2,400bp. The region with a power law in $P(\Delta)$ covers 39.98\% of the euchromatic binding sites. Furthermore, the chromosomal  average $\alpha$ of mouse values 0.3292$\pm$0.0068, covering 43.93\% of binding sites, and it is fitted in the interval 20bp$\leq\Delta\leq$ 4.1kbp.

For the region beyond polynomial fitting, $P(\Delta)$ decays exponentially, $P(\Delta)\approx e^{-\Delta/ \lambda}$. The characteristic length $\lambda$ for humans  values $\lambda=19.28\pm0.24$kbp and 15kbp $\leq\Delta \leq$ 99kbp is the exponential adjustment region, comprising 16.36\% of binding sites. In the case of the mouse, the genomic $\lambda$ values $18.06\pm0.29$kbp with 11kbp$\leq\Delta\leq$ 89kbp, containing 23.77\% of CTCFbs. 
 
A similar feature is described for the human K562 CTCF binding sites distribution \cite{Chen-2012}. However, we cannot compare the power law CTCFbs distribution for the entire genome directly with their cluster analysis \cite{Chen-2012}, since the power law has not a characteristic length by definition. Thus we use a cluster analysis, assuming those CTCFbs to be nearest neighbors that are within 3058bp and hence in one  particular cluster. We choose 3058bp because this is the median for the complete genome $\Delta$ as well as this is close to the upper limit of the power fitting, Table \ref{gbands2}. Thus, 63,68\% of our cluster-like structures can be classified as singletons (isolated CTCFbs), while \cite{Chen-2012} reports 38.94\%. The groups with 2, 3, 4, 5, 6 and more than 6 CTCFbs values respectively 18.09\%, 7.08\%, 3.45\%, 2.11\%, 1.26\% and 4.26\% while \cite{Chen-2012} indicate 25.09\%, 14.60\%, 8.79\%, 5.22\%, 3.10\% and 4.26\% in their cluster map. Although we have more singletons in our results, we have the same  percentage for cluster-like structures with more than 6 CTCFbs reported by \cite{Chen-2012}.

We do not restrict our analysis just to human and mouse. We confirm the existence of cluster-like structures in pig and goat, where we find 316,919 and 264,286 CTCFbs respectively. Both average chromosomal CTCFbs densities $<l_{\mbox{\scriptsize ctcf}}>$  are compatible with the human and mouse, but direct comparison should be avoided because we exclude the heterochromatin in the human genome. $\alpha$ values are 0.484$\pm$0.013 and 0.5409$\pm$0.0064 for pig and goat respectively. They contain 44.27\% (pig) and 41.31\% (goat) of the binding sites. Both species have the same regions for $\alpha$ fitting: 14bp $\leq\Delta\leq$ 2,000bp, but the domains for $\lambda$ adjustments are different: 14kbp $\leq\Delta\leq$ 99kbp for pig, covering 13.75\% of binding sites; and 18kbp $\leq\Delta\leq$ 99kbp in the case of goat, composing 13.31\% of CTCFbs. 

$P(\Delta)$ is not limited just to polynomial and exponential fittings. We have many CTCFbs that are  13 bp apart from  each other as well. 5.67\% of human euchromatin, 6.39\% of mouse without chromosome Y, 7.58\% of pig and 7.19\% of goat binding sites are in the region $0<\Delta\leq$ 13bp, and $P(\Delta)$ distributions are not uniform. We observe few binding sites with $\Delta$=2, 5 or 7 bp and the height of $P(\Delta)$ is species dependent.By the way, 0.23\%, 0.40\%, 0.49\% and 0.46\% of binding sites are $\Delta=0$ distance respectively in human, mouse, pig and goat, {\it i.e.} the CTCF has multiple  binding modes in these sites as mentioned previously. 

We also apply our method to the fruit fly and localize 8,962 binding sites. Although the genome size is just 5\% of mammals, $P(\Delta)$ of {\it Drosophila melanogaster} resembles mammal with a well-defined power law $\alpha=0.454\pm0.012$ and exponential decay $\lambda=18.08\pm 0.52$kbp. The polynomial and exponential fittings are along 14bp $\leq\Delta\leq$ 14,000bp and 15kbp $\leq\Delta\leq$ 78kbp, covering 62.73\% and 29.83\%. 5.87\% of the binding sites are 13bp or less distant each other and 0.12\% has $\Delta=0$. 

We study the genome of {\it A. aegypti} and identify 39,777 binding sites. We do not find a power law, but 16.33\% and 3.57\% of binding sites are respectively 144$\pm$5bp ($\Delta_1$) and 287$\pm$5bp ($\Delta_2$) at a distance of each other, Fig. \ref{fittings01}(e). We remark that we need 146bp to wrap one nucleosome. 0.37\% of sites has multiple binding modes. $P(0<\Delta\leq\mbox{13 bp})$ is unlike the other genomes, since  1.26\% of CTCFbs are just at one  bp distance of each other. When we consider a region of 4.5kbp $\leq\Delta\leq$ 99kbp for the exponential fitting, we have $\lambda=42.1\pm2.0$kbp. The exponential fitting contains 61.47\% of CTCFbs.

This odd behavior can be observed in mouse chromosome Y too, where we find 2,512 binding sites. The low density of $<l_{\mbox{\scriptsize ctcf}}>=32,742$bp per predicted CTCFbs hides a surprise. This $<l_{\mbox{\scriptsize ctcf}}>$ is just 9\% higher than {\it A. aegypti}, and there is neither a power law  nor an exponential decay. The number of binding sites in the region where $0<\Delta\leq$ 13bp is minimal, is just 0.6\%. We do not observe multiple CTCF binding modes, $P(\Delta=0)=0$. Although mouse chromosome Y lacks a power law and an exponential decay, 35.51\% of binding sites presents well-defined $\Delta$ distances:  0.96\%, 1.47\%, 0.92\%, 2.15\%, 6.21\%, 0.80\%, 1.83\%, 1.15\%, 5.02\%, 2.03\%, 2.87\%, 1.15\%, 2.83\%, 1.15\%, 2.03\% and 2.95\% of the binding sites are 208$\pm$5 ($\Delta_1$), 780$\pm$7 ($\Delta_2$), 931$\pm$1 ($\Delta_3$), 1,539$\pm$8 ($\Delta_4$), 1,927$\pm$2 ($\Delta_5$), 4,736$\pm$4 ($\Delta_6$), 6,949$\pm$11 ($\Delta_7$), 7,161$\pm$22 ($\Delta_8$), 7,587$\pm$12 ($\Delta_9$), 17,523$\pm$16 ($\Delta_{10}$), 20,862$\pm$21 ($\Delta_{11}$), 23,551$\pm$27 ($\Delta_{12}$), 28,153$\pm$29 ($\Delta_{13}$), 28,460$\pm$21 ($\Delta_{14}$), 31,452$\pm$25 ($\Delta_{15}$) and 40,565$\pm$91bp distance of each other ($\Delta_{16}$) in Fig. \ref{fittings01}(d), respectively. 

We test our method for {\it Plasmodium falciparum} (low unicellular eukaryote) and {\it Arabidopsis thaliana} (plant),  where CTCF is absent \cite{Heger-2012}. In the case of {\it P. falciparum}, Table \ref{gbands2}, the number of CTCFbs spotted by our method is so small  that we cannot even build $P(\Delta)$. As a matter of fact, there are only 3$\pm$3 CTCF binding sites per chromosome. We have better statistic for {\it A. thaliana}, Table \ref{gbands2}, where we detected 1,595 CTCFbs. The expected binding sites in the region $0<\Delta\leq$13bp is represented by 7.4\% of CTCFbs and they are at 6$\pm$4bp distance of each other. We do not report multiple binding modes for these species, $P(\Delta=0)=0$ and there is neither a polynomial nor an exponential decay for $P(\Delta)$. These binding sites detected by our method are false positives. They are born from the $P(S_i)$ statistics in $P_{\mbox{\scriptsize flank}}$ from pattern identification and other limitations outlined along this manuscript. 

{\it Caenorhabditis elegans} is another interesting specimen. Although this worm lost its CTCF gene along the evolution \cite{Heger-2009}, we encounter 2,086 binding sites, possible remains of its segmented body past \cite{Heger-2009}. In the region $0<\Delta\leq$13bp, we have 5.27\% of the binding sites and there are two sites with multiple binding modes. These values are compatible with mammalian genomes. But we do neither find a power law  nor an exponential decay in its CTCFbs distribution. The density of CTCFbs in {\it C. elegans} is $48.2\pm 7.7$kbp. This $<l_{\mbox{\scriptsize ctcf}}>$ is not far from human centromeric domains ($38 \pm7$ kbp per CTCFbs, Table \ref{gbands2}). Here we have $P(0<\Delta\leq 13\mbox{bp})$=3.3\%  and 0.2\% of sites present multiple binding mode, but we do neither find a power law nor an exponential decay. 

One may argue the absence of a power law and exponential decay in $P(\Delta)$ is due to the low density of CTCFbs $<l_{\mbox{\scriptsize ctcf}}>$ in the human centromeric domain or in  mouse chromosome Y. However, we have an unusual concentration of binding sites in the human noncentromic and nontelomeric heterochromatin regions (gvar). These domains are:  the entire 3q11.2 and 19q12; the initial part of 9q12, 19p12 and Yq12; final part of 1q12, 13p11.2, 16q11.2 and 22p11.2. They have 14Mbp of the length, represent 44.7\% of all heterochromatic CTCFbs and 0.3\% of the sites has multiple binding modes. Nevertheless, similarities with euchromatic segments end at this point. We do not observe the $P(0<\Delta\leq\mbox{13 bp})$ distribution of the mammal genomes, but 7.5\% of CTCFbs are 10$\pm$2bp distant each other. We do neither observe a power law nor an exponential decay in $P(\Delta)$ too. 

Finally, we report just 20 binding sites in the chromosome 4 of fruit fly. But, this number is too small for conclusive results.

\section{Discussion}

The molecular basis for the four letters alignment is the hydrogen bonds of the nucleotides. The adaptation of the Peyrad-Bishop model of the DNA melting for the transcription factor binding \cite{Zhu-2007} also considers the hydrogen bonds as responsible for the electronic pattern along the genome. Although the Peyrad-Bishop explains successfully the separation of the base pair under the temperature variation in polymerase chain reaction, transcription factors, as EGR1, SP1 and CTCF, do not open the double Helix in their search for binding sites. They scan the dDNA, inserting zinc fingers into the major and minor grooves of DNA and probing for $\pi$-orbital electronic patterns~\cite{Wolfe-1999}. So, the Peyrad-Bishop cannot be applied directly for the search of the transcription factor binding site. However, the nucleotide $\pi$-orbitals have successfully been described by the extended ladder model, which interprets dDNA as semiconductor-like material~\cite{Senthilkumar-2005,Mehrez-2005}. When we apply the extended ladder to transcription factor binding DNA sequences, patterns as in Fig.~\ref{tabela12}(c) appear. Again we emphasize that this semiconductor-like description is {\it in situ} condition dependent. 

The electronic nucleotide alignment using the extended ladder model considers the dDNA in atmosphere, low vacuum or Tris-HCl buffers  \cite{Oiwa-2016,Yoo-2001,Cai-2000,Taniguchi-2006}. There is no consensus about the electronic transport properties of dDNA, since the experimental frameworks change the electronic properties of DNA \cite{Taniguchi-2006}. EDTA or HEPES buffers may induce an  electric insulator effect ~\cite{Pablo-2000}. However, at room temperature and under tris(hydroxymethyl)-aminomethane and hydrochloride salt (Tris-HCl), a traditional physiological buffer with pH=7.382 at 37$^{\circ}$C \cite{Durst-1972}, dDNA has a semiconductor like behavior \cite{Oiwa-2016,Cai-2000,Taniguchi-2006}. This buffer may emulate the living HeLa cytosol and nucleus conditions, {\it i.e.} an aqueous solution with pH around 7.35 \cite{Llopis-1998}. Under this circumstance, we may adopt the charge transport formalism to the nucleotide analysis. 

The charge transport formalism adds a new layer over the nucleotide alignment. We are not  restricted just to 4 letter pattern. As in the three ZFs of EGR1 and SP1~\cite{Oiwa-2016}, CTCF also anchors ZFs in the nucleotides with the most stable electrons, {\it i.e.} the ground state and the lowest occupied $\pi$-orbitals. These nucleotides are at the core of the consensus motif. Numerical simulation in~\cite{Lei-2015} also show that the central ZFs are the most relevant nucleotide for the CTCF binding.

When we examine our CTCFbs with those in~\cite{Chen-2012}, we observe many mismatches. One source for the predict and experimental ChIP-seq CTCFbs differences is the sample number for nucleotide and electronic pattern in Fig.~\ref{tabela12} and for the statistics of $P_{\mbox{\scriptsize core}}$ and $P_{\mbox{\scriptsize flank}}$ in section pattern identification. The 23 samples do not cover all possibilities, although they catch the most common features. Actually,  \cite{Xie-2007,Chen-2012} also mention these additional motifs beyond the 20-mer consensus motif, positions -9 to 11 in Fig.~\ref{tabela12}(d). Moreover, we are not considering homologous CTCFs~\cite{Ding-2008}. The samples in Fig.~\ref{tabela12}(c) belong to mouse and human only, and we are defining one common CTCF pattern for them. Although we may expect a general mechanism from a common arrangement, we may foresee specie depend variations in the electronic pattern.

We also introduce noise when we consider the second criteria $\cup_iS_i$ for $S_{\mbox{\scriptsize flank}}$ in  pattern identification, based on the nucleotide occurrence, Fig.~\ref{tabela12}(e). This term  plays a similar role as the background frequency correction in DNA sequence motifs. Although, this approach adds flexibility, it introduces systematic error in site prediction: the method will consider some false motifs.

CTCF can bind to dDNA in multiple ways as in shown in Fig.~\ref{tabela12}(c), but we combined all binding possibilities in one simple binding pattern. Indeed the literature about CTCF motifs does not consider multiple CTCF binding possibilities. However, experimental results~\cite{Filippova-1996}, numerical simulations~\cite{Lei-2015} and careful charge analysis of the tips of the zinc fingers show many viable binding arrangements. Unfortunately, the sample number in this work is too small for each individual binding configuration. Thus, we joint all, following the literature~\cite{Chen-2012,Ong-2014}.

The process for positioning the CTCF binding sites in the K562 uses hg38, which is a consensus sequence of 9 healthy males \cite{GenBank}, while K562 is a tumoral cell from a woman~\cite{Lozzio-1975}. So, we are using  sequences of one person to find the position in the consensus of 9 others individuals. Most of sequences will be placed in the correct spot, but we expect discrepancies between these data. 

Despite all limitations and criticisms about our method and the ChIP-seq technique, we have 29.8$\pm$3.9\% of perfect matching and 20.2\% of near matching ($\|i_{\mbox{\scriptsize ctcf}}-i_c\|<474$bp, median, Q2), 25\% with intermediate misplacing ($474\leq\|i_{\mbox{\scriptsize ctcf}}-i_c\|<2,376$bp, third quartile, Q3) and 25\% of mismatching bigger than 2,374bp. Surprisingly, \cite{Marinov-2014} reports similar result: 55\% of successful identification, around 25\% with intermediate quality and 20\% with poor quality. \cite{Marinov-2014} attributes the poor quality data to the low depth reading in ChIP-seq assays. \cite{Chen-2012} also reports nearly 30\% of CTCFbs without the characteristic 20-mer consensus motif in ChIP-seq data and \cite{Kim-2007} reports the 20-mer motif in just over 75\% of experimentally identified CTCFbs. Moreover, using limited quality data from ENCODE and only 5 samples of K562 ubiquitous CTCF binding sites do not help us in the evaluation of the electronic nucleotide alignments. Nevertheless, extensive tested and analyzed genome using huge ENCODE data by independent peer as  \cite{Kim-2007,Chen-2012, Fu-2008} are rare. Otherwise, we may estimate the amount of misleading binding sites captured by our method from the {\it P. falciparum} and {\it A. thaliana}, Table \ref{gbands2}.
 
There is no CTCF gene for protozoan and plants~\cite{Heger-2012}. So, these binding sites are false positives generated by $P(S_i)$ statistics in $P_{\mbox{\scriptsize flank}}$ in pattern identification. Since we have around one CTCFbs in 9kbp for mammals (human, mouse, pig and goat), we estimate from 2\% to 13\% of false positives in our technique considering {\it P. falciparum} and {\it A. thaliana} as negative controls. {\it C. elegans} is not a good negative test. Although this worm lost its CTCF genes along its evolution \cite{Heger-2012}, this organism still hold CTCFbs.

There are three regions for the probability distribution $P(\Delta)$ of the distance $\Delta$ of two consecutive CTCF binding sites in human, mouse, pig, goat and fruit fly. In the first region, the binding sites appear in tandem and they are very close to each other, $0<\Delta\leq$ 13bp. The second region starts at 11bp $\sim$ 20bp and extends in between 2kbp to 17kbp. These are the domains for the power law fitting. The third  domain ranges from 2kbp $\sim$ 15kbp to 62kbp $\sim$ 99kbp, when we have an exponential decay in $P(\Delta)$. Beyond 100kbp, we have visible structures in optical microscope as the high packed chromatin, coordinated by scaffold proteins in mitotic cells. But, this very large scale organization is not a topic in this paper.

In the $0<\Delta\leq$ 13 bp domain, the number of binding sites represents 5.67\% to 7.58\% of the total. Further, there are always binding sites with multiple reading modes: 0.12\% $\leq P(\Delta=0)\leq$ 0.49\%. Here, we have multiple binding modes due to the molecular CTCF shape variations~\cite{Lei-2015,Filippova-1996}, beyond the different dDNA reading modes due to the symmetries of the genomic code. The upper limit of this region is delimited by the size of the CTCF binding site. The binding site from the position -11 to 13 in Fig.~\ref{tabela12}, resulting in a 24bp of length, is compatible with the literature, where the length values 11bp $\sim$ 60bp~\cite{Chen-2012,Ong-2014,Ohlsson-2001,Phillips-2009}. However, we need just 4$\sim$5 ZFs for the CTCF-DNA attachment, using just 13 nucleotides. So, it is  not surprise that this region end at 13bp.

We have a power law for $\Delta$ beyond 13bp. This domain ranges from 11bp $\sim$ 20bp to 2kbp $\sim$ 17kbp, covering between 39.98\% to 62.73\% of binding sites. For these distances, CTCF may interact with dDNA as well as other transcription factors due to the N and C-terminals. In human, they are respectively 150 and 265 long amino-acid sequences with distinct highly acid and basic domains~\cite{Klenova-1993,Zlatanova-2009}. Further, the electronic nucleotide alignment in Fig.~\ref{tabela12}(b) shows consistently the presence of LUMOs and HOMOs around a binding site, reinforcing such  a possibility. Although the SysZNF database provide insights about the molecular structures of the head and end of homologous CTCFs~\cite{Ding-2008}, detailed studies about  N and C terminals interaction with DNA are rare and vague, despite experimental results~\cite{Filippova-1996}. 

The CTCF alone is not able to explain the power law. {\it Aedes aegypti} genome gives us a cue about the CTCF organization in these regions. The characteristic distances of 144$\pm$4bp and 287$\pm$5bp in $P(\Delta)$, Table~\ref{gbands2} and Fig.~\ref{fittings01}(e), reflect the action of the nucleosomes in chromatin. We need 147bp to wrap one nucleosome core. Moreover, the mouse chromosome Y has a recognizable 208$\pm$5bp distance in $P(\Delta)$, indicating a nucleosome wrapping by 147 bp with linker of 61bp long. Indeed the  mouse chromosome Y distinct distances 780$\pm$7bp, 931$\pm$1bp, 1,539$\pm$ and 1,927$\pm$2bp, Fig.~\ref{fittings01}(d), can be also interpreted as a chromatin with respectively  4, 5, 8 and 10 nucleosomes attached in the dDNA with two CTCF in  the extremities. The CTCFs of these complexes may connect each other creating small DNA-loops. In the case of $\Delta$ ranging from 4,736$\pm4$bp to 40,565$\pm$91bp, Fig.~\ref{fittings01}(d), we have  from 25 to 219 nucleosomes between the binding sites. The presence of nucleosomes around CTCF binding sites is confirmed by~\cite{Chen-2012,Fu-2008} as well as in Fig.~\ref{fittings01}(a). 

The interaction of CTCFs and nucleosomes result in a solenoidal, zig-zag ribbon or other irregular chromatin structures with a polynomial decay in $P(\Delta)$. The distribution of CTCFbs will have a cluster-like appearance, Fig.~\ref{fittings01}(e), troubling ChIP-seq procedures~\cite{Marinov-2014,Park-2009}. Binding sites in tandem will bring ambiguities in motif alignments used in the ChIP-seq protocol too.

The distance between these cluster-like CTCFbs groups can be examined by the behavior of $P(\Delta)$, when  $\Delta$ ranges from 2kbp $\sim$ 15kbp to 62kbp $\sim$ 99kbp. $P(\Delta)$ becomes exponential, because the probability in finding the next CTCFbs after $\Delta$ nucleotides is $p (1-p)^{\Delta}$, where $p$ is the probability of occurrence of the CTCFbs. We can approximate this expression as $pe^{-p\Delta}$, since $p<<1$. So, we expect an exponential decay in the case of random distribution of CTCFbs. Calling $p=1/\lambda$, we observe an exponential behavior for $P(\Delta)$, when $\Delta$ is bigger than 2kbp$\sim$15kbp. 

We may illustrate the power law and the exponential decay of $P(\Delta)$ in the mouse HOXA gene cluster (cf. Fig.~\ref{fittings01}(b)).  
The distance between CTCFbs inside of a cluster-like group never exceed 3058bp and obeys a power law with $\alpha=0.3292\pm0.0068$ in mouse. Nonetheless, we have a distance around 17kbp between A4 (*) and C5$\vert$6 as well as before A13 and after A13, and $\lambda=19,79\pm0.31$kbp in Table~\ref{gbands2}.
 
The number of binding sites is not small in the exponential distances, ranging from 13.31\% to 29.83\%. In the case of {\it A. aegypti,} we have 61.47\%. The chromatin folding process in these distances cannot be explained just with CTCF and nucleosomes. Multiple different chromosome folding for these $\Delta$ distances is mediated by non-histone proteins as cohesin, Ying and Yang 1 (YY1) and others \cite{Ong-2014,Zlatanova-2009}.

Moreover, CTCF may skip many binding sites~\cite{Ong-2014}. Monte Carlo simulations show that the depletion of histones along the chromatin has influence over the folding process  \cite{Dame-2014}. This is illustrated in the putative cluster-like binding sites of the genes A10, A11 and A13, indicated by diamonds in Fig.~\ref{fittings01}(b), where the binding sites were overlooked by CTCF. The number of binding sites localized by ChIP-seq is usually a fraction of the expected ones, with a chromosomal average of just one  in 42$\pm$12 human ubiquitous euchromatic CTCF binding sites in the K562 cells.

Finally, we are working with incomplete data. So, direct comparison between species must be done carefully. Major efforts from the community must be done seeking for less fragmented complete sequences. 
When the number of contigs are large and the size is small, 
most of them are too short for computing distances between binding sites and the segment number is excessive for handling them individually. 
The procedures described in this article are not automated yet. So, the manipulation of thousands of contigs is not viable. Furthermore, the many gaps will add noise in the probability distribution $P(\Delta)$ of the distance $\Delta$ between two consecutive binding sites. In fact, most of genomes deposited in GenBank are excessively fragmented, even those organized in chromosomes. However, new sequences deposited in GenBank overcome such limitations. The recently reviewed genomes of pig and goat have few gaps (see material), opening new perspectives to unveil the chromosomal organization in the coming years.

The CCCTC transcription factor binding sites (CTCFbs) have a characteristic $\pi$-orbital nucleotide motif. Mobile electrons are absent in the core of CTCF binding regions, i.e. we do neither observe highest occupied molecular orbitals (HOMO) nor lowest unoccupied molecular orbitals (LUMO) between ZF3 to ZF5. The CTCF may miss ZF2 or ZF6 binding with DNA. But, it cannot miss both simultaneously. There are at least three different ways to CTCF attach to the DNA. Our nucleotide alignment match with those reported by ~\cite{Kim-2007,Bell-2000,Ong-2014}. 

We report 335,088 predicted CTCFbs in the whole human genome, using the electronic nucleotide alignment~\cite{Oiwa-2016}. When we compare our results with the ubiquitous K562 chromatin immunoprecipitation with massively parallel DNA sequencing data (ChIP-seq)~\cite{Chen-2012}, we have 29.8$\pm$3.8\% of matching. And, 75\% of mismatches are with less than 2,352 bp distance between the measured one and the predicted from our method. These 2kbp discrepancies are expected because we use reduced number of  experimental sequences for the search of our electronic pattern and the limitations of the extended ladder model. However, larger mismatches ($>$2kbp) are due to  ChIP-seq assay: insufficient depth of reading~\cite{Marinov-2014,Park-2009}; the absence of the 20-mer consensus motif in the ChIP-seq data~\cite{Kim-2007,Chen-2012} or even position of multiple CTCF motifs, each one related with one possible binding pattern.

 When we combine our predicted CTCFbs and nucleosome positions, we localize 15 nucleosomes flanking CTCFbs as expected in \cite{Chen-2012,Chen-2014,Fu-2008,Clarkson-2019}. Furthermore, the distribution of nucleosomes around CTCF reveal asymmetry, reflecting the N and C-terminous molecular differences.

Furthermore, we confirm the experimental results of~\cite{Narendra-2015} with our theoretical study, detecting all CTCFbs in the mouse HOXA cluster.

We have studied the genomes of {\it Mus musculus} (mouse), {\it Sus scrofa} (pig), {\it Capra hircus} (goat), {\it Drosophila melanogaster} (fruit fly) and {\it Aedes aegypti} (mosquito) finding 277,027, 316,919, 264,286, 8,982 and 39,777 CTCF binding sites respectively.  We also analyzed {\it Caenorhabditis elegans}, {\it Plasmodium falciparum} and {\it Arabidopsis thaliana} as negative controls. Since {\it C. elegans}, protozoans and plants have no CTCF gene, there are few binding sites as expected.

The CTCFbs distribution along whole genomes of studied mammals and insects, totalizing 11.77 billion nucleotides, may be described as follows: For distances between 11bp$\sim$20bp and 2kbp $\sim$ 17kbp, CTCFbs compose cluster-like groups, where the interval $\Delta$ between
two consecutive binding sites obeys a power law with a coefficient $\alpha$ varying from 0.3292$\pm$0.0068 (mouse)  to 0.5409$\pm$0.0064 (goat).   There is no power law for the {\it Aedes} genome, but 19,9\% of binding sites are at 144$\pm$4 and 287$\pm$5bp distance of each other. These cluster-like CTCFbs groups are separated with a typical distance between 18.08$\pm$0.52kbp (fruit fly) to 42.1$\pm$2.0kbp ({\it Aedes}).

%%%%%%%%%%%%%%%%%%%%%%%%%%%%%%%%%%%%%%%%%%%%%%%%%%
%
% ACKNOWLEDGMENTS
%
%%%%%%%%%%%%%%%%%%%%%%%%%%%%%%%%%%%%%%%%%%%%%%%%%%
\section*{Acknowledgments}

The authors wish to thank Lei Liu and Sujeet Kumar Mishra for the discussions about zinc fingers and CTCF. This work is supported by Conselho Nacional de  Desenvolvimento Tecnol\'ogico e Cient\'{i}fico (CNPq), process number 248589/2013, Brazil. We acknowledge financial support by Deutsche Forschungsgemeinschaft
within the funding programme Open Access Publishing, by the
Baden-W\"{u}rttemberg Ministry of Science, Research and the Arts and by
Heidelberg University. The authors also acknowledge support by the state of Baden-W\"{u}rttemberg through bwHPC
and the German Research Foundation (DFG) through grant INST 35/1134-1 FUGG. This work was partially funded by the Deutsche Forschungsgemeinschaft (DFG, German Research Foundation) under Germany's Excellence Strategy EXC 2181/1 - 390900948 (the Heidelberg STRUCTURES Excellence Cluster). Kunhe Li would like
to thank the Chinese Scholarship Council (CSC) for the scholarship.

\section*{Author contributions}

NNO, CEC and DWH developed the method of the nucleotide alignment using local electronic density of states.
NNO and DWH developed, implemented all software as well as carried out all analysis. 
NNO, KL and DWH contribute with the nucleosome positioning using iNPS. NNO, CEC and DWH wrote the manuscript. All
authors read and approved the final version of the manuscript.

\section*{Conflict of Interest}

The authors declare that they have no conflict of interest.

\section*{Data Availability}

All CTCF binding sites computed for this work are freely available in HeiData.

%%%%%%%%%%%%%%%%%%%%%%%%%%%%%%%%%%%%%%%%%%%%%%%%%%
%
% REFERENCES
%
%%%%%%%%%%%%%%%%%%%%%%%%%%%%%%%%%%%%%%%%%%%%%%%%%%


\begin{thebibliography}{}

\end{thebibliography}


\begin{thebibliography}{}

\bibitem{Klenova-1993} Klenova EM, Nicolas RH, Paterson HF, Carne AF, Heath CM, Goodwin GH, Neiman PE, Lobanenkov VV (1993) CTCF, a conserved nuclear factor required for optimal transcriptional activity of the chicken c-myc gene, is an 11-Zn-finger protein differentially expressed in multiple forms. Mol. Cell. Biol. 13, 7612-7624

\bibitem{Kim-2007} Kim TH, Abdullaev ZK, Smith AD, Ching KA, Loukinov DI, Green RD, Zhang MQ, Lobanenkov VV, Ren B (2007) Analysis of the Vertebrate Insulator Protein CTCF-Binding Sites in the Numan Genome. Cell 128, 1231-1245

\bibitem{Xie-2007} 
Xie X, Mikkelsen TS, Gnirke A, Lindblad-Toh K, Kellis M, Lander ES (2007) Systematic discovery of regulatory motifs in conserved regions of the human genome, including thousands of CTCF insulator sites. PNAS USA 104, 7145-7150

\bibitem{Chen-2012} Chen H, Tian Y, Shu W, Bo X, Wang S (2012) Comprehensive Identification and Annotation of cell Type-Specific and Ubiquitous CTCF-Binding Sites in the Human Genome. PLoS One 7, e41374.

\bibitem{Iuchi-2001} Iuchi S (2001) Three classes of C2H2 zinc finger proteins. Cell Mol. Life Sci. 58, 625-635

\bibitem{Ding-2008} Ding G, Lorenz P, Kreutzer M, Li Y,Thiesen HJ (2009) SysZNF: the C2H2 zinc finger gene database.
Nucleic Acids Res. 37, D267–D273

\bibitem{Wolfe-1999} Wolfe SA, Nekludova L, Pabo C (1999) DNA Recognition by Cys2His2 Zinc Finger Proteins. Annu. Rev. Biophys. Biomol. Struct. 3, 183-212

\bibitem{Klug-2010} Klug A (2010) The Discovery of Zinc Fingers and Their Applications in Gene Regulation and Genome Manipulation. Annu. Rev. Biochem. 79, 213-31

\bibitem{Marinov-2014} Marinov GK, Kundaje A, Park PJ, Wold BJ (2014) Large Scale Quality Analysis of Published ChIP-seq Data G3. Genes, Genomes, Genetics 4, 209-223

\bibitem{Oiwa-2016} Oiwa NN, Cordeiro CE, Heermann DW (2016) The Electronic Behavior of
Zinc-Finger Protein Binding Sites in the Context of the DNA Extended
Ladder Model. Frontiers in Physics 4, 13/1-10

\bibitem{Bell-2000} Bell AC, Feisenfeld G (2000) methylation of a CTCF-dependent boundary controls imprinted expression of the Igf2 gene. Nature 405, 482

\bibitem{Essien-2009} Essien K, Vigneau S, Apreleva S, Singh LN, Bartolomei MS, Hannehalli S (2009) CTCF binding site classes exhibit distinct evolutionary, genomic, epigenomic and transcriptomic features. Genome Biology 10, R131

\bibitem{Ong-2014} 
Ong CT, Corces VG (2014)
CTCF: an architectural protein bridging genome topology and function. Nature Review Genetics 15, 234-246

\bibitem{encode-2007a} The ENCODE Project Consortium (2007) Identification and analysis of functional elements in 1\% of the human genome by the ENCODE pilot project. Nature 447, 799-816

\bibitem{GEOmn} Edgar R, Domrachev M, Lash AE (2002)
Gene Expression Omnibus: NCBI gene expression and hybridization array data repository.
Nucleic Acids Res. 30, 207-10 

\bibitem{Chen-2014} Chen W, Liu Y, Zhu S, Green CD, Wei G, Han J-D J (2014) Improved nucleosome-positioning algorithm iNPS for accurate nucleosome positioning from sequencing data. Nature Communications 5, 4909

\bibitem{NucMap} Zhao Y, Wang J, Liang F, Liu Y, Wang Q, Zhang H, Jiang M, Zhang Z, Zhao W, Bao Y, Zhang Z, Wu J, Asmann YW, Li R, Xiao J (2019) NucMap: a database of genome-wide nucleosome positioning map across species. Nucleic Acids Res. 47, D163–D169

\bibitem{Narendra-2015} Narendra V, Rocha PP, An D, Raviram R, Skok JA, Mazzoni EO, Reinberg D (2015) CTCF establishes discrete functional chromatin domains at the Hox clusters during differentiation. Science 347, 1017-1021


%%%%% begin samples 

%\bibitem{Filippova-2001} Filippova GN, Thienes CP, Penn BH, Cho DH, Hu YJ, Moore JM, Klesert TR, Lobanenkov VV, Tapscott SJ. CTCF-biniding sites flank CTG/CAG repeats and form a methylation-sensitive insulator at the DM1 locus. Nature Genetics 2001; 28: 335-343.

%\bibitem{Renaud-2007} Renaud S, Loukinov D, Abdullaev Z, Guilleret I, Bosman FT, Lobanenkov V, Benhattar J. Dual role of DNA methylation inside and outside of CTCF-binding regions in the transcriptional regulation of the telomerase hTERT gene. Nucleic Acids Res. 2007; 35: 1245-1256.

%\bibitem{Velazquez-2007} Rosa-Vel\'azquez IA, Rinc\'on-Arano H, Ben\'{i}tez-Bribiesca L. Epigenetic Regulation of the Human Retinoblastoma Tumor Suppressor Gene Promoter by CTCF. Cancer Res. 2007; 67: 2577-2585.

%\bibitem{Vostrov-1997} Vostrov AA, Quischke WW. The Zinc Finger Protein CTCF binds to the APB$\beta$ Domain of the Amyloid $\beta$-Protein Precursor Promoter. J. Bio. Chemistry 1997; 272: 33353-33359.

%\bibitem{Gombert-2009} Gombert WM, Krumm A. Targeted Deletion of Multip-le CTCF-Binding Elements in the Human C-MYC Gene Revels a Requirement for CTCF in C-MYC Expression. PLoS One 2009; 4: e109.

%\bibitem{Butcher-2004} Butcher DT, Mancini-DiNardo DN, Archer TK, Rodenhiser DI. DNA binding sites for putative methylation boundaries in the unmethylated region of the BRCA1 promoter. Int. J. Cancer 2004; 111: 669-678.

%\bibitem{Wylie-2000} Wylie AA, Murphy SK, Orton TC, Jirtle RL. Novel Imprinted DLK1/GTL2 Domain on Human Chromosome 14 Contains Motifs that Mimic Those Implicated in IGF2/H19 Regulation. Genome Research 2000; 10: 1711-1718.

%\bibitem{Saitoh-2000} Saitoh  N, Bell AC, Recillas-Targa F, West AG, Simpson M, Pikaart M, Felsenfeld G. Structural and functional conservation at the boundaries of the chicken $\beta$-globin domain. The EMBO Journal 2000; 19: 2315-2322.

%\bibitem{Hark-2000} Hark AT, Schoenherr CJ, Katz DJ, Ingram RS, Levorse JM, Tilghman SM. CTCF mediates methylation-sensitive enhancer-blocking activity at the H19/Igf2 locus. Nature 2000; 405: 486.

%%%%% end samples

\bibitem{GenBank} Benson DA, Cavanaugh M, Clark K, Karsch-Mizrachi I, Lipman DJ, Ostell J, Sayers EW (2013) GenBank. Nucleic Acids Res. 41, D36-42

\bibitem{Zhu-2007}
Zhu JX, Rasmussen KO, Balatsky AV, Bishop AR (2007) Local electronic structure in the Peyrard-Bishop-Holstein model. J. Phys.: Condens. Matter 19, 136203. 

\bibitem{Senthilkumar-2005} Senthilkumar K, Grozema FC, Guerra CF, Bickelhaupt FM, Lewis FD, Berlin YA, Ratner MA, Siebbeles LDA (2005) Absolute Rates of Hole Transfer in DNA. J. Am. Chem. Soc. 127, 14894-14903

\bibitem{Mehrez-2005} Mehrez H, Anantram MP (2005) Interbase electronic coupling for transport through DNA. Physical Review B71, 115405

\bibitem{Yoo-2001} Yoo KH, Ha DH, Lee JO, Park JW, Kim J, Kim JJ, Lee HY, Kawai T, Choi HY (2001) Electrical Conduction through Poly(dA)-Poly(dT) and Poly(dG)-Poly(dC) DNA Molecules. Physical  Review Letters 87, 198102-1

\bibitem{Dreos-2013} Dreos R, Ambrosini G, P\'erier RC, P. Bucher P (2013) EPD and EPDnew, high-quality promoter resources in the next-generation sequencing era. Nucleic Acids Res. 41, D157

\bibitem{Miller-1985} Miller J, McLachlan AD, Klug A (1985) Repetitive zinc-binding domains in the protein transcription factor IIIA from {\it Xenopus} oocytes. The EMBO Journal 4, 1609-1614

\bibitem{Nolte-1998} Nolte RT, Conlin RM, Harrison SC, Brown RS (1998) Differing roles for zinc fingers in DNA recognition: Structure of a six-finger transcription factor IIIA complex. PNAS USA 95, 2938-2943

\bibitem{Lei-2015} Liu L, Heermann DW (2015) The Interaction of DNA with multi-Cys2His2 zinc finger proteins. J. Phys.: Condens. Matter 27, 064107

\bibitem{Schneider-1986} Schneider TD, Storno GD, Gold L (1986) Information Content of Binding Sites on Nucleotide Sequences. J. Mol. Biol. 188, 415-431

\bibitem{Schneider-1990} Schneider TD, Stephens RM (1990) Sequence logos: a new way to display consensus sequences. Nucleic Acids Res. 18, 6079-6100

\bibitem{Bailey-1994} Bailey TL, Elkan C (1994) Fitting a mixture model by expectation maximization to discover motifs in biopolymers. Proceedings of the Second International Conference on Intelligent Systems for Molecular Biology 1994, 28-36

\bibitem{haeseleer-2006} D'haeseleer P (2006) What are DNA sequences motifs? Nature Biotechnology 24, 423-425

\bibitem{Setubal-1997} Setubal J, Meidanis J (1997) Introduction to computational molecular biology. PWS Publ., Boston, USA

\bibitem{Mount-2004} Mount DW (2004) Bioinformatics Sequence and Genome Analysis, 2nd Ed, Cold Spring Harbor, New York, USA

\bibitem{Zhong-2014} Zhong J, Wasson T, Hartemink AJ (2014) Learning protein-DNA interaction landscapes by integrating experimental data through computational models. Bioinformatics 30, 2868-2874

\bibitem{encode-2007b} Gerstein MB, Bruce C, Rozowsky JS, Zheng D, Du J, Jan O. Korbel JO, Emanuelsson O, Zhang ZD, Weissman S, Snyder M (2007) What is a gene, post-ENCODE? History and updated definition. Genome Res. 17, 669-681

\bibitem{Wing-1980} Wing R, Drew H, Takano T, Broka C, Tanaka S, Itakura K, Dickerson RE (1980) Crystal structure analysis of a complete turn of B-DNA. Nature 287, 755-758

\bibitem{Fu-2008} Fu Y, Sinha M, Peterson CL, Weng Z (2008) The Insulator Binding Protein CTCF Positions 20 Nucleosomes around Its Binding Sites across the Human Genome. PLoS Genetics 4, e1000138/1-13

\bibitem{GENCODE}Frankish A et al. (2019) GENCODE reference annotation for the human and mouse genomes.
Nucleic Acids Res. 47, D766-D773

\bibitem{UCSCbrowser} Kent WJ, Sugnet CW, Furey TS, Roskin KM, Pringle TH, Zahler AM, Haussler D (2002) The human genome browser at UCSC. Genome Res. 12, 996-1006

\bibitem{Segal-2006} Segal  E, Fondufe-Mittendorf Y, Chen L, Thåström A, Field Y, K Moore IK, Wang JPZ, Widom J (2006)
A genomic code for nucleosome positioning.
Nature 442, 772-8

\bibitem{Clarkson-2019} 
Clarkson CT, Deeks EA, Samarista R, Mamayusupova H, Zhurkin VB, Teif VB (2019)
CTCF-dependent chromatin boundaries formed by asymmetric nucleosome arrays with decreased linker length.
Nucleic Acids Res. 47, 11181–11196

\bibitem{Park-2009} Park PJ (2009) ChIP-seq: advantages and challenges of a maturing technology. Nature Reviews 10, 669-680

\bibitem{Heger-2012} Heger P, Marin B, Bartkuhn M, Schierenberg E, Wiehe T (2012)
The chromatin insulator CTCF and the emergence of methazoan diversity.
PNAS USA 109:,17507-17512 

\bibitem{Heger-2009} Heger P, Marin B, Schierenberg E (2009)
Loss of the insulator protein CTCF during nematode evolution.
BMC Molecular Biology 10, 84/1-14

\bibitem{Cai-2000} Cai L, Tabata H, Kawai T (2000) Self-assembled DNA networks and their electrical conductivity. Appl. Phys. Lett. 77, 3105

\bibitem{Taniguchi-2006} Tanigushi M, Kawai T (2006) DNA electronics. Physica E 33, 1-12

\bibitem{Pablo-2000} de Pablo PJ, Moreno-Herrero F, Colchero J, Herrero JG, Herrero P, Bar{\'o} AM, Ordej{\'o}n P, Soler JM, Artacho E (2000) Absence of dc-Conductivity in $\lambda$-DNA. Phys. Rev. Lett. 85, 4992

\bibitem{Durst-1972} Durst RA, Staples BR (1972) Tris/Tris.HCl: A Standard Buffer for Use in the Physiologic pH Range. Clinical Chemistry 18, 206-293

\bibitem{Llopis-1998} Llopis J, McCaffery JM, Miyawaki A, Farquhar MG, Tsien RY (1998) Measurement of cytosolic, mitochondrial, and Goldi pH in single living cells with green fluorescent protein. PNAS USA 95, 6803-6808

\bibitem{Filippova-1996} 
Filippova GN, Fagerlie S, Klenova EM, Myers C, Dehner Y, Goodwin G, Neiman PE, Collins SJ, Lobanenkov VV (1996)
An exceptionally conserved transcriptional repressor, CTCF, employs different combinations of zinc fingers to bind diverged promoter sequences of avian and mammalian c-myc oncogenes.
Mol. Cell. Biol. 16, 2802-13

\bibitem{Lozzio-1975} Lozzio CB, Lozzio BB (1975) Human chronic myelogenous leukemia cell-line with positive Philadelphia chromosome. Blood 45, 321–334

\bibitem{Ohlsson-2001} Ohlsson R. Renkawitz R, Lobanenkov V (2001) CTCF is a uniquely versatile transcription regulator linked to epigenetics and disease. Trends in Genetics 17, 520-527

\bibitem{Phillips-2009} Phillips JE, Corces VG (2009) CTCF: Master Weaver of the Genome. Cell 137, 1194-1211

\bibitem{Zlatanova-2009} Zlatanova J, Caiafa P (2009) CTCF and its protein partners: divide and rule? J. Cell Sci. 122, 1275-1284

\bibitem{Dame-2014} Tark-Dame M, Jerabek H, Manders EMM, Heermann DW, van Driel R (214) Depletion of the Chromatin Looping Proteins CTCF and Cohesin Causes Chromatin Compaction: Insight into Chromatin Folding by Polymer Modelling. PLOS Computational Biology 10, e1003877

%\bibitem{HeidelbergDB} {\color {red} *** http://... we need a DB ***}

\end{thebibliography}
\end{document}

%\linenumbers

% %%%%%%%%%%%%%%%%%%%%%%%%%%%%%%%%%%%%%%%%%%%%%%%%%
%
% FRONT MATTER
%
% %%%%%%%%%%%%%%%%%%%%%%%%%%%%%%%%%%%%%%%%%%%%%%%%%

%\title{A Comparative Prediction and Analysis of CTCF Binding Sites based on a First Principle Approach}

\begin{flushleft}
{\Large
\textbf\newline{Supplementary material: Prediction and Comparative Analysis of CTCF Binding Sites based on a First Principle Approach}
}
\newline

Nestor Norio Oiwa\textsuperscript{1,2},
Kunhe Li\textsuperscript{2},
Claudette El\'\i sea Cordeiro\textsuperscript{3},
Dieter W. Heermann\textsuperscript{2,*},

\bigskip
\textbf{1} Department of Basic Science, Universidade Federal Fluminense, Rua Doutor S\'{i}lvio Henrique Braune 22, Centro, 28625-650 Nova Friburgo,  Brazil
\\
\textbf{2} Institute for Theoretical Physics, Heidelberg University, Philosophenweg 19, D-69120 Heidelberg, Germany
\\
\textbf{3} Department of Physics, Universidade Federal Fluminense, Avenida Atl\^antica s/n, Gragoat\'a, 24210-346 Niter\'oi, Brazil
\\
\bigskip

%* nestoroiwa@if.uff.br,nestoroiwa@vm.uff.br
* heermann@tphys.uni-heidelberg.de

\end{flushleft}
%\author{Nestor Norio Oiwa}
%\email{nestoroiwa@if.uff.br}
%\affiliation{Department of Basic Science, Universidade Federal Fluminense, Rua Doutor S\'{i}lvio Henrique Braune 22, Centro, 28625-650 Nova Friburgo,  Brazil}
%\affiliation{Institute for Theoretical Physics, Heidelberg University, Philosophenweg 19, D-69120 Heidelberg, Germany}

%\author{Claudette El\'\i sea Cordeiro}
%\email{clau@if.uff.br}
%\affiliation{Department of Physics, Universidade Federal Fluminense, Avenida Atl\^antica s/n, Gragoat\'a, 24210-346 Niter\'oi, Brazil}

%\author{Dieter W. Heermann}
%\email{heermann@tphys.uni-heidelberg.de}
%\affiliation{Institute for Theoretical Physics, Heidelberg University, Philosophenweg 19, D-69120 Heidelberg, Germany}

\date{\today}

%%%%%%%%%%%%%%%%%%%%%%%%%%%%%%%%%%%%%%%%%%%%%%%%%%
%
% ABSTRACT
%
%%%%%%%%%%%%%%%%%%%%%%%%%%%%%%%%%%%%%%%%%%%%%%%%%%

%\keywords{CTCF binding sites, Cys2His2 zinc finger, extended ladder model}
Keywords: CTCF binding sites, Cys2His2 zinc finger, extended ladder model.

%\maketitle

\section*{S1: Selection of Well-known CTCF binding sites for electronic nucleotide alignment}

The first sequence in Table 1 is a file 2,139 bp long covering 
between genes DMPK and SIX5, related with Myotonic Dystrophy (DM)~\cite{Filippova-2001}. We can easily localize 
DM1 and DM2 CTCF binding sites, because they flank the repeated sequence $(CTG)_n$. The authors apply
gel mobility shift assay for CTCF-biding site identification.
The next file is the 1,161 bp long around the beginning of the first exon 
of the gene telomerase reverse transcriptase (TERT)~\cite{Renaud-2007}. This CTCF binding site is identified by 
ChIP, electrophoretic mobility shift (EMSA) and transient transfection assays 
We study the CTCF binding site at human retinoblastoma gene promoter~\cite{Velazquez-2007} using a fasta 700bp long file. The existence of this particular binding site is confirmed by EMSA in HeLa.
EMSA in HeLa cells are also used for
the binding site confirmation in the promoter of amyloid $\beta$-protein precursor (APP) gene~\cite{Vostrov-1997}. Here, we select a
2,149 bp nucleotide sequence for APP.
The CTCF binding sites of the v-myc avian myelocytomatosis viral oncogene homolog (MYC)
are identified by
ChIP assay~\cite{Gombert-2009}. We use a fasta file with a length of 1,366bp around the CTCF binding sites a and n.
The existence of CTCF sites in the breast cancer 1 (BRCA1) gene are confirmed using EMSA and
ChIP~\cite{Butcher-2004}. We target the same region form the reference map using a sequence 2,310 bp long surround CTCF1 and CTCF2 binding sites. 
We also take the 2,870 bp long 
maternally expressed imprinted gene 3 (MEG3) between DLK1 and GTL2 genes. This is a putative CTCF binding sites, similar to H19 and Igf2 domains, validated by methylation assay \cite{Wylie-2000}.
In the $\beta$-globin (HBE) CTCF binding site, validated by EMSA, we consider a sequence with 1330 bp  flanking the folate receptor 1 gene~\cite{Saitoh-2000}.
Finally, we have the H19/Insulin-like grown factor 2 gene (Igf2) CTCF binding site clusters for mouse and human.
In the case of Mus musculus, we are using h1 to h5 cluster~\cite{Hark-2000}.
We select a 3,430 bp long nucleotide sequence around 3kbp upstream of H19.
The h4 binding site of this cluster is particularly interesting because this putative binding site, spotted using traditional nucleotide alignment, is not confirmed experimentally.
We take a fasta file with 550 bp with h1 to h7 for human. The methylation of this cluster has already studied experimentally using EMSA~\cite{Bell-2000}.
We are not using the original sequences in our work, but regions around the mentioned CTCF binding sites in the reference map. 
We apply BLAST for finding the sequences of the interest \cite{BLAST}.

\section*{S2: Extended ladder model \cite{Oiwa-2016}}

Since a $L\times L$ matrix with billion size $L$ is not possible for the eigenvalue and eigenvector computation, we slipt the complete genome with $L$ nucleotides in windows with length n=200bp. Then we compute the local density of states 
from the eigenvalues $E_k$ and eigenvectors $\phi_i^k, k=1,...,n_e,$ of the nucleotide in position $i$ using the extended ladder model. Here $n_e$ is the number of electrons in the double helix (dDNA) and we have $2n$ nucleotides with $n$ base pairs. 

The model consider one double DNA chain with $n$ base pairs, totaling $2n$ nucleotides.
Actually our model does not consider nucleotides, but {\bf nucleosides},
i.e. the {\bf nucleotide} with the phosphate group. But, we simplify the nomenclature calling nucleosides by nucleotides. 
The spinless free electron of
the nucleotide $\pi$-orbital is described by~\cite{Oiwa-2016,Zhu-2007},
\begin{equation}
 H=H_e+H_{eb}+H_b. 
      \label{modelo02}
\end{equation}
Here $H_e$  is the electronic degree of freedom without nucleotide coupling,
\begin{eqnarray}
H_e&=&\sum_{i=1}^{2n}\epsilon_iC_i^{\dagger}C_i
     + ( \sum_{i=1}^{n-1}t_{2i-1,2i+1}C_{2i-1}^{\dagger}C_{2i+1} \
     +\sum_{i=1}^{n-1}t_{2i,2i+2}C_{2i}^{\dagger}C_{2i+2} \nonumber\\
     &&+\sum_{i=1}^{n-1}t_{2i-1,2i}C_{2i-1}^{\dagger}C_{2i}
     +\sum_{i=1}^{n-1}t_{2i-2,2i+1}C_{2i-2}^{\dagger}C_{2i+1} )+H.c.\\ \nonumber\label{He01}
\end{eqnarray}
where $C^{\dagger}_i$ and $C_i$ are the electron creation and
annihilation operators at site $i$, $\epsilon_i$ is the on-site ionization
energy, $n$ is the number of nucleotides and $t_{ij}$ is the electron
hopping rate between nucleotides $i$ and $j$. The lattice 
considered In Eq.~\ref{He01} is the extended ladder and
the electronic hopping rates in $H_e$ are the same in the literature~\cite{Oiwa-2016,Senthilkumar-2005,Mehrez-2005,Sarmento-2009,Zilly-2010}. 
Moreover, $H_{eb}$ represents the coupling between the free electron and the nucleotide displacement field,
\begin{equation}
    H_{eb}=\alpha_v\sum_{i=1}^{2n}y_iC_i^{\dagger}C_i\label{Heb01}
\end{equation}
where $y_i$ is the displacement of the electronic cloud from the equilibrium in the nucleotide. $H_{eb}$ controls the gap size
between HOMO and LUMO and we fix $\alpha_v=1.0$. In this way, the gap in our spectra will be in accordance with those reported in literature~\cite{Senthilkumar-2005,Mehrez-2005,Sarmento-2009,Zilly-2010,Shapir-2008,Wang-2004}. 
Finally, $H_b$ is the interaction of the electron with the nucleotide:
\begin{equation}
   H_b=\sum^{2n}_{i=1}[D_i ( e^{-a_iy_i}-1)^2+\frac{k_v}{2}(y_i-y_{i-1})^2],\label{Morse01}\\
\end{equation}
where $D_i$ and $a_i$ are parameters of the Morse potential, $k_v$ is
the spring constant of the anharmonic interaction between two contiguous
base-pairs. Concerning the parameters for the Morse potential, we are using those extensively suggested in the density functional
literature: $D_A$, $D_T$, $D_C$ and $D_G$ are respectively 0.25eV, 0.44eV, 0.33eV and 0.45eV~\cite{Richardson-2004,Gu-2012};
$a_A$, $a_T$, $a_C$ and $a_G$  are correspondingly  3.0\AA$^{-1}$, 3.0\AA$^{-1}$, 3.0\AA$^{-1}$
and 2.5\AA$^{-1}$ \cite{Chen-2007,Chen-2009}; and $k_v= 0.0125$eV \cite{Oiwa-2016}.

We study the electronic part $H_e$ and $H_{eb}$ of the Hamiltonian in
Eq.~\ref{modelo02} computing the eigenvalue $E_k$ and eigenvectors
$\phi_i^k$, $i,k=1,...,2n$, of the $2n\times 2n$ Hermitian matrix $H_e+H_{eb}$~\cite{Oiwa-2016}.
Given an initial $\{y_i\}$, we diagonalize $H_e+H_{eb}$ calculating the electronic
occupation in each site $<n_i>$, where $n_i=\sum_{k=1}^{n_e}|\phi_i^{k}|^2$ and $n_e$ is the number of electrons
in the system. This set of $<n_i>$ will be used for the $y_i$ estimate in the Langevin equation, given by
\begin{equation}
<\frac{\partial H_b}{\partial y_i}+\frac{\partial H_{eb}}{\partial y_i}>=0,\label{selfconsist01}
\end{equation}
where $<...>$ is the average over the free electrons in the system. 
We update the values of $\{y_i\}$, using fourth-order Runger-Kutta method in the Langevin
equation. The new $\{y_i\}$ set is inserted again in the matrix $H_e+H_{eb}$. 
We repeat the iteration until we achieve the minimum
local adiabatic electronic and structural configuration. 
Since we wish to analyze massive amount of data, we rewrite the code in R used in~\cite{Oiwa-2016} to C++, increasing the performance over the original program by factor of a thousand. The iteration method for solving Eqs. $H_e+H_{eb}$ and the self consistent Eq.~\ref{selfconsist01} have already been described in~\cite{Oiwa-2016,Zhu-2007}. 

\begin{figure}[th]
   \centering
   \includegraphics[width=0.8\linewidth]{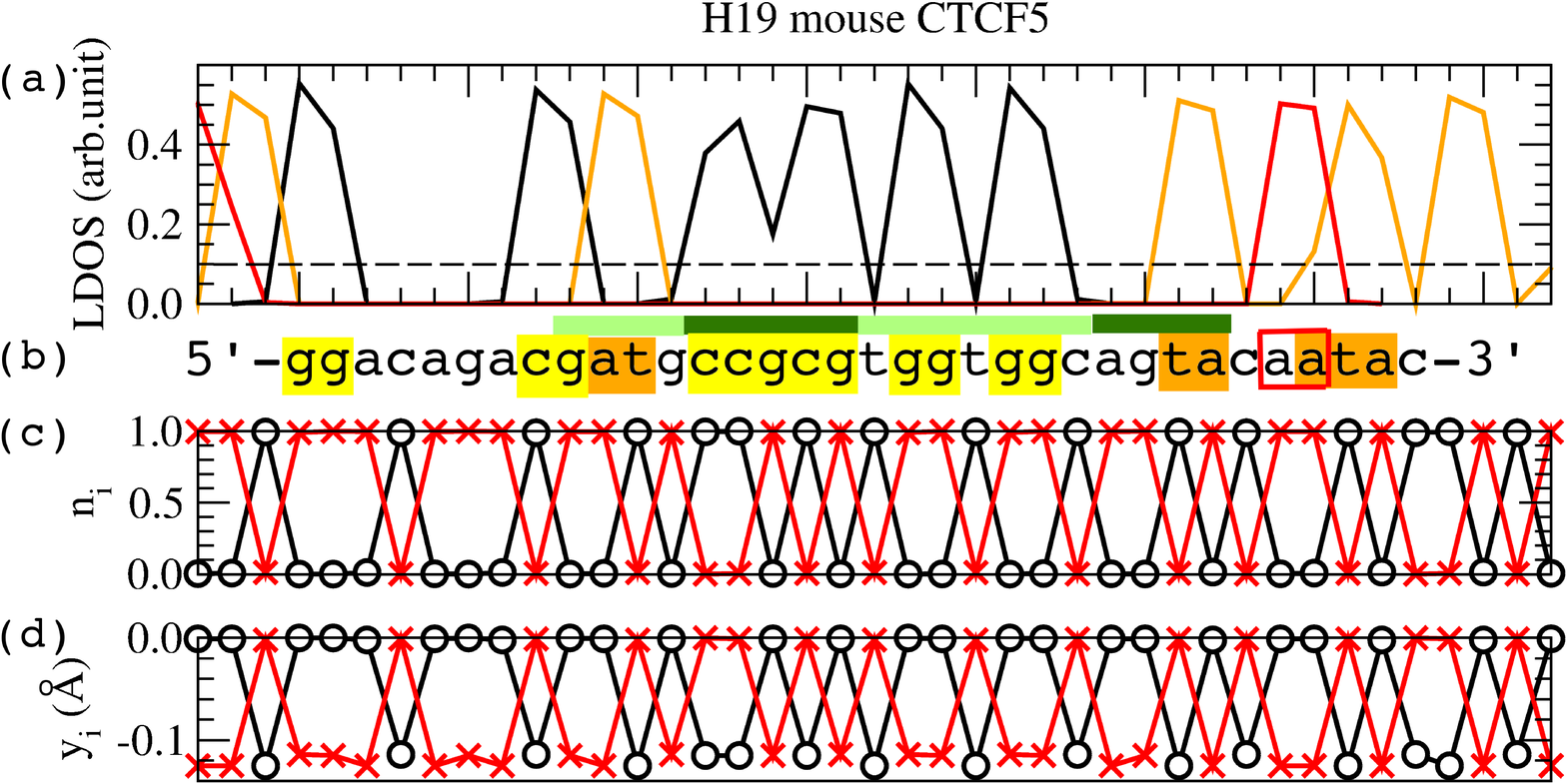}
%     \epsfxsize =\textwidth
%     \begin{center}
%          \leavevmode
%          \epsfbox{H19mousectcf5a.eps}
%     \end{center} 
     \caption{(a) is the ground state (black), HOMO (orange) and LUMO  (red) of the local electronic density of states (LDOS) for H19 Mouse CTCF5. The quota of 10\% used in (b) is in dashed line. 
(b) is nucleotide sequence of the CTCF5-DNA binding site in reverse complementary strand. Nucleotides with at least 10\% of probability in finding one ground state, HOMO and LUMO electrons are indicated respectively in yellow, orange and red bordered boxes.
We have  in light and dark green the four modulus in \cite{Ong-2014}.
    (c) is  the probability for finding one electron in the direct strand (black)
     and the complementary strand (red), when the valence band is
     completely filled, $n_e=n$. (d) is the field displacements
     $y_i$ in the Morse potential with $n_e=n$ for the direct strand
     (black) and for the complementary strand (red).}
     \label{H19mousectcf5}
\end{figure}

Using the results for SP1 and EGR1 in our previous work \cite{Oiwa-2016}, we define electrons with a maximum 8.02eV of the energy as bottom of the molecular orbital. In this work we call them ground states in order to simplify their understand in the context of the paper, since they include the ground states. We call lowest unoccupied molecular orbital (LUMO) those electrons with $9.1\leq E_k\leq 9.4$eV, and highest occupied molecular orbital (HOMO) are electrons with $8.52\leq E_k\leq 8.60$eV. We show a typical result for the H19 mouse CTCF 5 in Fig. \ref{H19mousectcf5}. The local density of states (LDOS) of the ground states is in black lines in Fig. \ref{H19mousectcf5}(a), and HOMO are in orange and the LUMO electrons are in red. Once we estimate the shape of the electronic cloud along the DNA chain, the nucleotides with at least 10\% of probability in finding ground state, HOMO or LUMO electrons are marked respectively in yellow, orange or red bordered boxes (c.f. Fig. \ref{H19mousectcf5}(b)). Assuming that the valence band is completed filled and the conduction band is empty, $n_e=n$, we usually have 100\% of probability $n_i$ in finding electron in cytosine and thymine (pyrimidines), Fig. \ref{H19mousectcf5}(c), as we reported in our previous article \cite{Oiwa-2016}. Finally, we may distinguish the different nucleotides too \cite{Oiwa-2016}, because guanine and cytosine have a displacement field $y_i$ around $-0.11$\AA, while adenine and thymine have -0.12\AA \hspace{0.1cm} in Fig. \ref{H19mousectcf5}(d). The displacement field $y_i$ is the rearrangement of the $\pi$-orbital of nucleotide $i$ in function of electron-base interaction \cite{Oiwa-2016,Zhu-2007}.

\section*{S3: Selected Genomes}

% Complete human genome assemble hg38

We apply the method for the 24 {\it Homo sapiens} chromosomes (GRCh38/hg38) \cite{GenBank}. 
Although the human genome was drafted in 2001 \cite{Lander-2001}, the numerous gaps remains due to repetitive domains. 
The assemble hg38 still has 303 contiguous sequences (contigs), instead of 24 assembled molecules. 
The statistics of the fragmentation of human genome is N50=56,413,054bp and L50=19, where contigs with length N50 or longer include half of the bases of the molecular chromosomal assembly
and L50 is the number of contigs  that contains half of base pairs \cite{GenBank}.
Since most of 303 contigs are small and restricted in particular regions, we consider only those bigger than 1 million of base pairs (1 Mbp), diminishing the amount in 96. 
The contigs have also 
small gaps with few base pairs of length, filled with N or another letter. Although
the statistics of N50 and L50 is provided by \cite{GenBank}, the real genomic fragmentation must be checked before, since these small gaps are frequently neglected in L50 and N50. We admit a maximum of 10 small gaps per 1Mbp and the sum of the small gaps should be smaller than 1kbp as the acceptable contiguous sequence.
Despite these exclusion criteria, we still cover around 91.8\% of the 3,088,269,837 bp long complete genome (column $L$ in Tab. 1). Our genome length account is smaller than \cite{GenBank}, because we consider the assembly molecule, excluding unlocalized scaffolds.

The 21 chromosomes of {\it Mus musculus} (mouse, build GRCm38.p6) are also studied. This genome is 2,725,521,371 long and the contigs cover 96.1\%. This genome has N50=32,273,079bp and L50=26 \cite{GenBank}. We reduce the 353 contigs to 159 using the same criteria for human contigs.

%Description of other GenBank files
%====================

Although GenBank holds genetic information of thousands of species \cite{GenBank},
most of reference genomes are still very fragmented as we will discuss later. However, in 2016 pig and goat became available  with acceptable N50 and L50 statistics, {\it i.e.} N50 bigger than 1Mbp and L50 smaller than 100. Beyond this values, the genome is too fragmented and not practical.

The 19 chromosomes of {\it Sus scrofa} (pig, breed Duroc, build Sscrofa11.1) 
have N50= 48,231,277bp and L50=15 \cite{GenBank}. This is a 2,435,262,063bp genome with 98.1\% of coverage. We do not study the chromosome Y, because it is too fragmented.  

The 29 chromosomes of {\it Capra hircus} (goat, build ARS1, breed San Clemente, N50= 26,244,591bp, L50=32) are 
2,466,191,353bp long and cover 84.3\% of genome. Our length is shorter than those reported by \cite{GenBank}, because we do not consider the chromosome X due to its excessive fragmentation and 
there is not data about chromosome Y in ARS1. 

We do not restrict our CTCF analysis to mammals. \cite{Heger-2012} reports CTCF in insects too. 
We consider the 6 chromosomes of 
{\it Drosophila melanogaster} (fruit fly, Release 6, N50=19,478,218bp and L50=3) \cite{GenBank}. We exclude the chromosome Y, since it is divided in too many segments. The 133,880,608bp long genome has 96.0\% of covering.
In the case of the 3 chromosomes of 
{\it Aedes aegypti} (build AaegL5.0, N50=11,758,062bp and L50=30), the genome is 1,195,030,408
bp long, covering entire genome  \cite{GenBank}. This is the mosquito responsible for the transmission of yellow fever, denge, zika and chinguya. 
Again the problems of fragmentation of the genomes do not allow us to advance beyond the mentioned insect genomes.

At the end, we apply the extended ladder model for some negative tests.

{\it Caenohabditis elegans} is a worm with 6 chromosomes that lost its CTCF gene along the evolution \cite{Heger-2012,Ong-2014}.
Since there are no gaps in the sequence, the evaluation of N50 and L50 is meaningless for this genome.
We use the build WS262, a 100,272,607bp long assembly \cite{GenBank}. 

We study the genome of {\it Plasmodium falciparum}, build ASM276v2. This is the protozoan with 14 chromosomes, which causes malaria. The 23,264,338 long genome has not gene for CTCF.

We analyze the genome of {\it Arabidopsis thaliana} (buid TAIR10, N50=11,194,537bp and L50=5)  \cite{GenBank}. 
The 5 chromosomes are 119,146,138bp long and have 97.4\% of coverage.
Although \cite{GenBank} announce {\it A. thaliana} as complete, there are many gaps filled by Y (pyrimidine) or other letters.
Since plants have no CTCF, we do not expect them in this genome.

After working with the many GenBank files enlisted above, we conclude that only genomes with N50 bigger than 1Mbp and L50 smaller than 100 are viable for CTCFbs analysis proposed in this paper.

\section*{S4: Example of $P_{\mbox{\scriptsize core}}$ and $P_{\mbox{\scriptsize flank}}$ estimate}

As an illustration of the pattern identification method, consider 5'-aa cc gg ccg cg agg ttg cag tg ca-3'.

The subsequence 5'-agg tgg cag-3' belongs to the core zinc fingers $\{ZF5, ZF4, ZF3\}$. In the case of first triplet agg, we have 9 nucleotide sequences for the motif agg in the column ZF5 along the 23 selected CTCFbs in Fig. 1(c). We mark this motif as agg(9) in Fig. 1(f). Then, we associate a probability of $P(S_{\mbox{\scriptsize ZF5}})=\frac{9}{23}$, when the sequence agg appear along the genome. Same procedure is made for $P(S_{\mbox{\scriptsize ZF3}} )=\frac{6}{23}$ and $P(S_{\mbox{\scriptsize ZF4}})=\frac{12}{23}$, resulting $P_{\mbox{\scriptsize core}}=37,62\%$. This probability indicates a valid CTCFbs, because $P_{\mbox{\scriptsize core}}\geq 9.0\%$.

The subsequences 5'-aa cc gg ccg cg-3' and 5'-tg ca-3' are the flanking sequences. They are associated with $\{ZF9, ZF8a, ZF8b, ZF7, ZF6, ZF2a, ZF2b\}$. When we consider only the motifs associated, we have $P(S_{\mbox{\scriptsize ZF9}})=\frac{6}{23}$, $P(S_{\mbox{\scriptsize ZF8a}})=\frac{3}{23}$, $P(S_{\mbox{\scriptsize ZF8b}})=\frac{8}{23}$, ... ,
$P(S_{\mbox{\scriptsize ZF2b}})=\frac{7}{23}$, resulting in $\left[\prod_kP(S_k)\right]^{1/7}=33,19\%$ in $P_{\mbox{\scriptsize flank}}$. We also compute 
$P(a_{\mbox{\scriptsize -11}})=\frac{11}{23}$,
$P(a_{\mbox{\scriptsize -10}})=\frac{9}{23}$,
$P(c_{\mbox{\scriptsize -9}})=\frac{7}{23}$,
... ,
$P(g_{\mbox{\scriptsize -1}})=\frac{13}{23}$,
$P(t_{\mbox{\scriptsize 10}})=\frac{13}{23}$,
$P(g_{\mbox{\scriptsize 11}})=\frac{14}{23}$,
$P(c_{\mbox{\scriptsize 12}})=\frac{11}{23}$,
$P(a_{\mbox{\scriptsize 13}})=\frac{11}{23}$
and calculate the geometric average of the nucleotide occurrence $\left[\prod_i P(S_i)\right]^{1/15}=52.72\%$. So, 
$P_{\mbox{\scriptsize flank}}=42,96\%$, which is a putative CTCFbs, since it is bigger than 6.5\%. 

Since both $P_{\mbox{\scriptsize core}}$ and $P_{\mbox{\scriptsize flank}}$ are valid CTCFbs, the sequence 5'-aa cc gg ccg cg agg ttg cag tg ca-3' is a good candidate for the CTCFbs.

\section*{S5: K562 ChIP-seq data}

% K562

We verify our electronic pattern using ChIP-seq data of the K562 cells, deposited at The Encyclopedia of DNA Elements (ENCODE). K562 is an immortal cell strain that come from a 53 year woman \cite{Lozzio-1975} and ENCODE is a databank seeking the integration of the many biological functions along the genomes \cite{encode-2007a,encode-2007b}. We use the following K562 files: ENCFF002CEL, ENCFF002CLS, ENCFF002CLT, ENCFF002CWL and ENCFF002DDJ with respectively 51,992, 45,603, 11,533, 54,387 and 43,247 CTCFbs each one. Since they are GRCCh37 build (hg19) and we consider GCCh38 assembly coverage  (hg38), we apply the NCBI Remapping Service available in \cite{remapping}, converting hg19 to hg38 assemble. Only the ubiquitous binding sites in ChIP-seq data are used, because the ChIP-seq technology is not mature with possible false sites as we discuss along the paper. 
We localize 61,254 binding sites for K562 cells, of which 8,786 are ubiquitous. Since we are using updated ENCODE files, these values are different from \cite{Chen-2012}, where they found 67,986 CTCFbs with 19,036 ubiquitous. 5,817 sites of the ChIP-seq data are in the negative G-bands, representing 66.2\% of the total. The bands with the 25\% and 50\% of Giemsa stain responses have respectively 1122 and 987, resulting in 12.8\% and 11.2\% of the experimental data. The darker bands with quota 75\% and 100\% have 547 (6.2\%) and 298 ChIP-seq binding sites in K562 cells (3.4\%), respectively. And we report 15 binding sites (0,2\%) in the heterochromatic domains.

\section*{S6: NucMap}

%The NucMap, National Genomics Data Center, China, is a public database specialized in nucleosome positioning for multiple species \cite{NucMap}. 
In the usual nucleosome positioning method, the distribution profile of nucleosome positioning come from micrococal nuclease digestion with high-throughput sequencing data (MNase-seq) \cite{Zhang-2008}. After denoising, inflection points are detected in this profile, using Laplacian of Gaussian Convolution. Then the nucleosome positions are estimated from the region delimited by theses inflection points as maximums or minimums. The improved nucleosome-positioning algorithm (iNPS) increase the number of detected nucleosomes, considering derivatives of Gaussian convolution too \cite{Chen-2014}. 

%NucMap has a good collection of such high quality nucleosome positioning data.

%%%%%%%%%%%%%%%%%%%%%%%%%%%%%%%%%%%%%%%%%%%%%%%%%%
%
% ACKNOWLEDGMENTS
%
%%%%%%%%%%%%%%%%%%%%%%%%%%%%%%%%%%%%%%%%%%%%%%%%%%
%\section*{Acknowledgments}

%The authors wish to thank Lei Liu {\color{blue} and Sujeet Kumar Mishra} for the discussions about zinc fingers and CTCF. This work is supported by Conselho Nacional de  Desenvolvimento Tecnol\'ogico e Cient\'{i}fico (CNPq), process number 248589/2013, Brazil. We acknowledge financial support by Deutsche Forschungsgemeinschaft within the funding programme Open Access Publishing, by the Baden-W\"{u}rttemberg Ministry of Science, Research and the Arts and by Heidelberg University. The authors also acknowledge support by the state of Baden-W\"{u}rttemberg through bwHPC and the German Research Foundation (DFG) through grant INST 35/1134-1 FUGG.

%%%%%%%%%%%%%%%%%%%%%%%%%%%%%%%%%%%%%%%%%%%%%%%%%%
%
% REFERENCES
%
%%%%%%%%%%%%%%%%%%%%%%%%%%%%%%%%%%%%%%%%%%%%%%%%%%